\documentclass[conference]{IEEEtran}

\usepackage{graphicx}
\usepackage{balance}  
\usepackage[linesnumbered,vlined,ruled]{algorithm2e}
\usepackage{lipsum}
\usepackage{amsmath}
\usepackage{amssymb}
\usepackage{physics}
\usepackage{subfigure}
\usepackage{tabularx}
\usepackage{color}
\usepackage{hyperref}
\usepackage{float}
\usepackage{listings}
\usepackage{multirow}
\usepackage[normalem]{ulem}
 \usepackage{longtable}
\usepackage{enumitem}
\usepackage{multirow}
\usepackage{booktabs}
\usepackage{capt-of}
\usepackage{pifont}
\usepackage{ulem}
\usepackage{soul}
\usepackage{cancel}
\usepackage{bm}
\usepackage{url}
\usepackage{caption}
\def\BibTeX{{\rm B\kern-.05em{\sc i\kern-.025em b}\kern-.08em
    T\kern-.1667em\lower.7ex\hbox{E}\kern-.125emX}}

 \graphicspath{{./Graph/}, {./Fig/}, {./Legend/}}

\DeclareRobustCommand*\cal{\mathcal}

\newtheorem{example}{Example}
\newtheorem{lemma}{Lemma}

\newcommand{\kw}[1]{{\ensuremath {\mathsf{#1}}}\xspace}

\newcommand{\stitle}[1]{\vspace{0.2ex} \noindent{\textbf{#1}}}
\newcommand{\mc}[1]{{\ensuremath {\mathcal{#1}}}\xspace}

\newcommand{\attrs}[1]{\kw{attrs}(#1)} 
\newcommand{\ord}{\kw{ord}} 
\newcommand{\pos}[1]{\kw{pos}(#1)} 
\newcommand{\dom}[1]{\kw{dom}(#1)} 

\newcommand{\R}{R\xspace} 
\newcommand{\A}{A\xspace} 
\newcommand{\Rs}{\mc{R}} 
\newcommand{\q}{Q\xspace} 
\newcommand{\D}{D\xspace} 
\renewcommand{\H}{H\xspace} 
\newcommand{\T}{{\cal T\xspace}} 

\newcommand{\Q}{\q} 

\newcommand{\Qs}{\mc{Q}} 

\newcommand{\p}{p\xspace} 

\newcommand{\ti}[1]{t^{#1}} 
\newcommand{\extension}[2]{\kw{val}(#1 \rightarrow #2)} 

\newcommand{\ghd}{\kw{GHD}} 
\newcommand{\apj}{\kw{ADJ}} 
\newcommand{\hcube}{\kw{HCube}} 
\newcommand{\hcubej}{\kw{HCubeJ}} 
\def\gj{\kw{GHD Join}} 
\def\lf{\kw{Leapfrog}} 

\usepackage{mathtools}

\long\def\comment#1{}

\def\AS{\kw{AS}}
\def\LJ{\kw{LJ}}
\def\OK{\kw{OK}}
\def\WB{\kw{WB}}
\def\WT{\kw{WT}}
\def\EN{\kw{EN}}

\begin{document}


\title{Fast Distributed Complex Join Processing}


\author{
    \IEEEauthorblockN{Hao Zhang\IEEEauthorrefmark{1}, Miao Qiao\IEEEauthorrefmark{2}, Jeffrey Xu Yu\IEEEauthorrefmark{1}, Hong Cheng\IEEEauthorrefmark{1}}
    \IEEEauthorblockA{\IEEEauthorrefmark{1}The Chinese University of Hong Kong
    \\\{hzhang, yu, hcheng\}@se.cuhk.edu.hk}
    \IEEEauthorblockA{\IEEEauthorrefmark{2}The University of Auckland
    \\\{miao.qiao\}@auckland.ac.nz}
}

%
%
%
%



\maketitle
\begin{abstract}
	Big data analytics often requires processing complex join queries in parallel in distributed systems such as Hadoop, Spark, Flink. The previous works consider that the main bottleneck of processing complex join queries is the communication cost incurred by shuffling of intermediate results, and propose a way to cut down such shuffling cost to zero by a one-round multi-way join algorithm. The one-round multi-way join algorithm is built on a one-round communication optimal algorithm for data shuffling over servers and a worst-case optimal computation algorithm for sequential join evaluation on each server. The previous works focus on optimizing the communication bottleneck, while neglecting the fact that the query could be computationally intensive. With the communication cost being well optimized, the computation cost may become a bottleneck. To reduce the computation bottleneck, a way is to trade computation with communication via pre-computing some partial results, but it can make communication or pre-computing becomes the bottleneck. With one of the three costs being considered at a time, the combined lowest cost may not be achieved. Thus the question left unanswered is how much should be traded such that the combined cost of computation, communication, and pre-computing is minimal.
	
	In this work, we study the problem of co-optimize communication, pre-computing, and computation cost in one-round multi-way join evaluation. We propose a multi-way join approach \apj (Adaptive Distributed Join) for complex join which finds one optimal query plan to process by exploring cost-effective partial results in terms of the trade-off between pre-computing, communication, and computation.We analyze the input relations for a given join query and find one optimal over a set of query plans in some specific form, with high-quality cost estimation by sampling. Our extensive experiments confirm that \apj outperforms the existing multi-way join methods by up to orders of magnitude. 
\end{abstract}

\section{Introduction}

\comment{Join queries processing is one of the important issues in query processing, and join queries over relations based on the equality on the common attributes are commonly used in many real applications. To overcome the load of processing large and complex join queries, it requires to process queries on distributed platforms, such as data-flow engines, e.g., Hadoop, Spark, Flink, which are widely adopted due to their efficiency and scalability. Such systems are well optimized for OLAP tasks that deal with acyclic joins for large datasets in parallel by utilizing a series of binary join, which joins two relations at a time. However, they are not optimized for complex join queries in terms of the number of relations involved in a join query, which can be with cycles if a join is represented as an edge between two relations to be joined. Such complex join queries have been used in the analysis of the characteristic of the complex network \cite{yaveroglu_revealing_2014} and the query of knowledge graph \cite{chu_theory_2015}. The reason they fail to process such queries effectively is that complex join queries tend to produce huge intermediate results to be shuffled among servers when they are processed in a distributed system over a cluster of servers, which causes communication bottleneck. The inefficiency roots from the fact that the servers need to exchange the intermediate results in multi-rounds when they execute a join query by a sequence of distributed binary joins.}

Join query processing is one of the important issues in query processing, and join queries over relations based on the equality on the common attributes are commonly used in many real applications. Large-scale data analytics engines such as Spark \cite{armbrust_spark_2015}, Flink \cite{carbone2015apache}, Hive \cite{thusoo2009hive}, F1 \cite{shute_f1:_2013}, Myria \cite{wang2017myria}, use massive parallelism in order to enable efficient query processing on large data sets. Recently, data analytics engines are used beyond traditional OLAP queries that usually consist of star-joins with aggregates. Such new kind of workloads \cite{nam2020sprinter} contain complex FK-FK joins, where multiple large tables are joined, or where the query graph has cycles, and has seen many applications, such as querying knowledge graph \cite{elliott2009complete}, finding triangle and other complex patterns in graphs \cite{ammar_distributed_2018}, analyzing local topology around each node in graphs, which serves as powerful discriminative features for statistical relational learning tasks for link prediction, relational classification, and recommendation \cite{liu2013social, rossi2012transforming}.

\begin{figure}
	[tbp]
	
	\begin{tabular}
		[t]{c} \subfigure[One-Round Vs Multi-Round] { 
		\includegraphics[width=0.45\columnwidth]{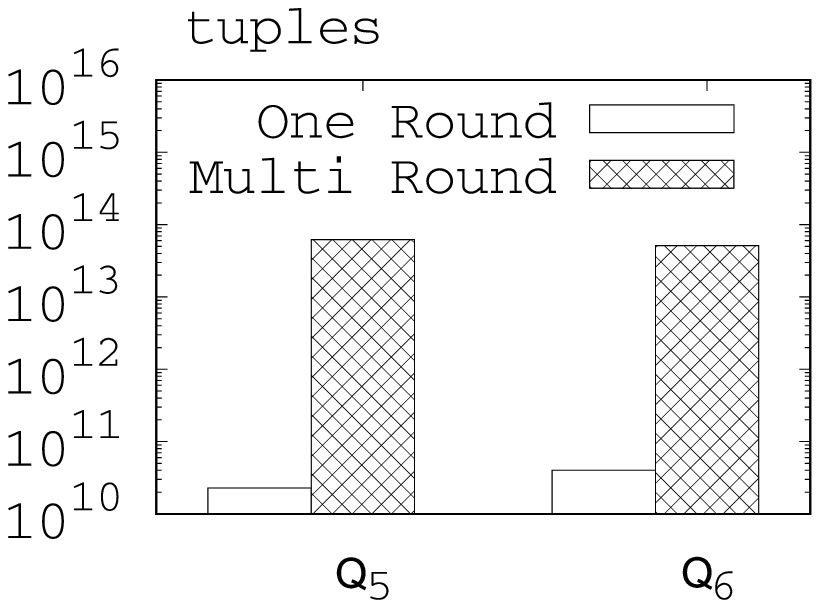} }
		
		\subfigure[Comm-First Vs Co-Opt] { 
		\includegraphics[width=0.45\columnwidth]{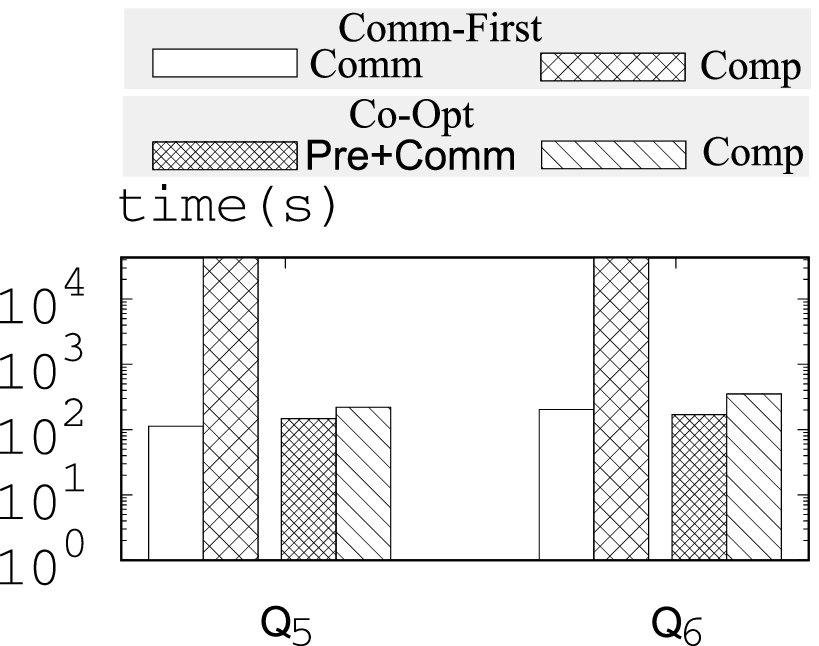} } 
	\end{tabular}

	\caption{Comparisons using two join queries, $Q_5$ and $Q_6$ (refer to Sec.\ref{sec:Exp_Setup}), over the \LJ dataset (refer to Table~\ref{tab:dataset})\label{fig:intro_example}. Here, ``Comm" denotes communication cost, ``Comp" denotes computation cost, ``Pre+Comm" denotes pre-computing cost plus computation cost.}
	
	\vspace{-0.2cm} 
\end{figure}

However, data analytics engines process complex joins by decomposing them into smaller join queries, and combining intermediate relations in multiple rounds, which suffers from expensive shuffling of intermediate results. To address such inefficiency, one-round multi-way join \hcubej is proposed \cite{chu_theory_2015}, which requires no shuffling after the initial data exchange. The one-round multi-way join processes a join query in two stages, namely, data shuffling and join processing. In the data shuffling stage, \hcubej shuffles the input relations by an optimal one-round data shuffling method \hcube~\cite{afrati_optimizing_2011, beame_communication_2013}. In the join processing stage, \hcubej uses an in-memory sequential algorithm \lf~\cite{veldhuizen_triejoin:_2014} at each server to join the data received. It can be seen in Fig.~\ref{fig:intro_example}(a) that the one-round multi-way join outperforms the multi-round binary join significantly, regarding the number of shuffled tuples, for complex join queries.

\comment{ A comparison of the total numbers of tuples transmitted between the one-round method and multi-round method is shown in. For both queries, we can see that the numbers of tuples shuffled by the one-round method are significantly lower as no tuples of intermediate results are shuffled between multiple rounds. }

However, the one-round multi-way join algorithm has a deficiency, since it puts communication cost at a higher priority to reduce than the computation cost by considering the communication cost as the dominating factor, which is not always true. The main reason is that the computation of complex multi-way join can be inherently difficult. We tested the communication-first strategy of \hcubej in our prototype system using optimized \hcube for data shuffling and \lf for join processing. As shown in the first two bars for each of the two queries ($Q_5$ and $Q_6$ in Sec.\ref{sec:Exp_Setup}) in Fig.~\ref{fig:intro_example}(b), the communication cost can be small, but the computational cost can be high. Overall, the performance may not be the best as expected.

In this paper, we study how to reduce the total cost by introducing pre-computed partial results with communication, computation, and pre-computing cost being considered at the same time. As shown in Fig.~\ref{fig:intro_example}(b), by our approach, we can reduce the computation cost significantly with some additional overhead for communication and pre-computing cost. This problem is challenging since we may cause one cost larger when we reduce the other cost, and the search space of potential pre-computed partial results is huge. The main contributions are given as follows. 

\begin{itemize}
	\item We identify the performance issue of processing $\q$ using \hcubej due to the unbalance between computation and communication cost, and propose a simple mechanism to trade computation cost with communication and pre-computing cost such that the total cost is reduced for a multi-way join query $\q$.
	\item We study how to effectively find cost-effective pre-computed partial results from overwhelmingly large search space, and join them and the rest of relations in an optimal order. To find such an optimal query plan, we reduce the search space of query plans to filter ineffective query plans early, and propose a heuristic approach to explore cost-effective pre-computed partial results and join orders.
	\item We propose a simple yet effective distributed sampling process with a theoretic guarantee to provide accurate cardinality estimation for query optimization. 
	\item We implement a prototype system \apj and propose several implementation optimizations that significantly improve the performance of \hcube, reduce the storage cost, and eliminate some redundant computation of \hcubej.
	\item We conducted extensive performance studies, and confirm that our approach can be orders of magnitude faster than the previous approaches in terms of the total cost. 
\end{itemize}

The paper is organized as follows. We give the preliminary of this work, and discuss \hcube, \lf algorithms, and the main issues we study in this work in Section~\ref{sec:preliminary}. We outline our approach in Section~\ref{sec:adj}, and discuss how to perform cardinality estimation via distributed sampling in Section~\ref{sec:p_estimation}. In Sec~\ref{sec:hcubeopt}, we discuss the implementation optimization of our prototype system. Section~\ref{sec:relate}, we discuss the related work, and in Section~\ref{sec:Experiment} we report our experimental studies. We conclude our work in Section~\ref{sec:conclude}. 

\section{Preliminaries} \label{sec:preliminary}

A database $D$ is a collection of relations. Here, a relation $R$ with schema $\{A_1, A_2, \cdots, A_n\}$ is a set of tuples, $(a_1, a_2, \cdots, a_n)$, where $a_i$ is a value taken from the domain of an attribute $A_i$, denoted as $\dom{A_i}$, for $1 \leq i \leq n$.Below, we use $\attrs{R}$ to denote the schema (the set of attributes) of $\R$. A relation $R$ with the schema of $\attrs{R}$ is a subset of the Cartesian product of $\dom{A_1} \times \dom{A_2} \times \cdots \times \dom{A_n}$ for $A_i \in \attrs{R}$. We focus on natural join queries (or simply join queries). A natural join query, $\q$, is defined over a set of $m$ relations, $\Rs = \{R_1, R_2, \cdots, R_m\}$, for $m \geq 2$, in the form of
\begin{equation}
	\q(\attrs{\q}) ~{}\mbox{:-}~{}R_1(\attrs{R_1}) \bowtie \cdots \bowtie R_m(\attrs{R_m}). 
\label{eq:join} \end{equation}

Here, the schema of $\q$, denoted as $\attrs{\q}$, is the union of the schemas in $\Rs$ such as $\attrs{\q} = \cup_{R_i \in \Rs} \attrs{R_i}$. For simplicity, we assume there is an arbitrary order among the attributes of $\Q$, denoted as $\ord$, and $A_i$ denotes the i-th attribute in $\ord$. We also use $\Rs(\q)$ to denote the set of relations in $\q$. A resulting tuple of $\q$ is a tuple, $\tau$, if there exists a non-empty tuple $t_i$ in $R_i$, for every $R_i \in \Rs$, such that the projection of $\tau$ on $\attrs{R_i}$ is equal to $t_i$ (i.e., $\Pi_{\attrs{R_i}} \tau = t_i$). The result of a join $\q$ is a relation that contains all such resulting tuples. A join query $\q$ over $m$ relations $\Rs$ can be represented as a hypergraph $\H = (V, E)$, where $V$ and $E$ are the set of hypernodes and the set of hyperedges, respectively, for $V$ to represent the attributes of $\attrs{\q}$ and for $E$ to represent the $m$ schemas. As an example, consider the following join query \q over five relations,
\begin{equation}
	\begin{split}
		\q(a, b, c, d, e)~{}\mbox{:-}~{}& \R_{1}(a, b, c) \bowtie \R_{2}(a, d) \bowtie \R_{3} (c, d) \bowtie \\
		& \R_{4}(b, e) \bowtie \R_{5}(c, e) 
	\end{split}
	\label{eq:qexample} \end{equation}

\begin{figure}
	[tbp] \center 
	\includegraphics[width=0.95\columnwidth]{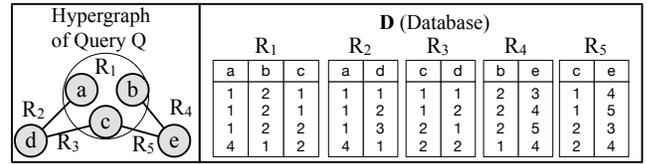}
	\caption{The hypergraph of query $\q$ (Eq~(\ref{eq:qexample})), and an example of database $\D$} \vspace{-0.2cm} 
\label{fig:Running_Example_1} \end{figure}

Its hypergraph representation $\H$ is shown in Fig.~\ref{fig:Running_Example_1} together with the 5 relations. Here, $V = \attrs{\q} = \{a, b, c, d, e\}$, and $E = \{e_1, e_2, e_3, e_4, e_5\}$ for $e_1 = \attrs{R_1}$, $e_2 = \attrs{R_2}$, $e_3 = \attrs{R_3}$, $e_4 = \attrs{R_4}$, and $e_5 = \attrs{R_5}$. In the following, we also use $V(\H)$ and $E(\H)$ to denote the set of hypernodes and the set of hyperedges for a hypergraph $\H$.

\begin{figure}
	[tbp] \center 
	\includegraphics[width=0.95\columnwidth]{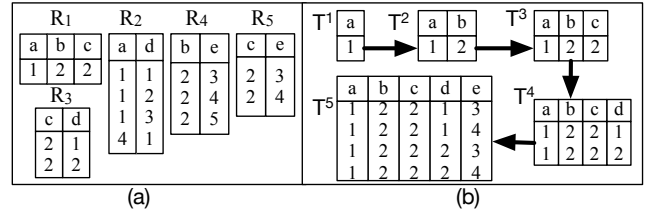}
	\caption{(a) The tuples shuffled to server $S_0$ with hypercube of coordinate $(0, 0, 0, 0, 0)$. (b) \lf at the server $S_0$ with hypercube of coordinate $(0, 0, 0, 0, 0)$} \vspace{-0.5cm} 
\label{fig:Running_Example_2_1} \end{figure}

\subsection{\lf and \hcube Join Algorithms} \label{sec:hcjoin}

We discuss \hcubej~\cite{chu_theory_2015} to compute join queries in a distributed system over a cluster of servers, where the database $D$ is maintained at the servers disjointly. \hcubej is built on two algorithms, namely, \hcube~\cite{afrati_optimizing_2011,beame_communication_2013} and \lf~\cite{veldhuizen_triejoin:_2014}, where \hcube is a one-round communication optimal shuffling method that shuffles data to every server in the cluster, and \lf is a fast in-memory sequential multi-way join algorithm to process the join query at each server over the data shuffled to it. For a join query $\q$ over $m$ relations, $\Rs = \{R_1, R_2, \cdots, R_m\}$, \hcube is proven in theory to be the optimal method in worst-case sense for transmitting the tuples to servers such that each server can evaluate the query on its own without further data exchange. \lf~\cite{veldhuizen_triejoin:_2014} is proven in theory to be the optimal method in worst-case sense to evaluate a join query $\q$, while binary join could be sub-optimal. Also, \lf is an iterator-based algorithm, which leaves little footprint in memory when processing the query.

\stitle{LeapFrog Join}~\cite{veldhuizen_triejoin:_2014} is one of the state-of-the-art sequential join algorithms for a join query $\q$ over $m$ relations, $\Rs = \{R_1, R_2, \cdots,$ $ R_m\}$~(Eq.~(\ref{eq:join})). Let $\attrs{\q}$ be the schema of $\q$ for $n = |\attrs{\q}|$. \lf is designed to evaluate $\q$ based on the attribute order $\ord$ using iterators. Let $t^i$ be an $i$-tuple that has $i$ attributes of $A_1$ to $A_{i}$. The \lf algorithm is to find the $t^{i+1}$ tuples by joining the tuple $t^i$ with an additional $A_{i+1}$ value recursively until it finds all $n$ attribute values for $\q$.

The \lf algorithm is illustrated in Algorithm~\ref{algo:leapfrog} for a given-input $i$-tuple $t^i$. The initial call of \lf is with an empty input tuple $t^0$. Below, we explain the algorithm assuming that the input is a non-empty $i$-tuple, $t^i$, for $i > 1$. Let $\Rs_{i+1}$ be the set of relations $R$ in $\q$ if $R$ contains the $(i$+$1)$-th attribute $A_{i+1}$ in order such as $\Rs_{i+1} = \{\R~{}|~{}\A_{i+1} \in \R~{}\mbox{and $\R$ is a relation appearing in $\q$}\}$ (line~4). To find all $A_{i+1}$ values that can join the input $i$-tuple $t^i$, denoted as $\extension{\ti{i}}{\A_{i+1}}$, (line~5), it is done as follows. Here, for simplicity and without loss of generality, we assume $\Rs_{i+1} = \{R, R'\}$. First, for $R$, let $As$ be all the attributes that appear in both $\attrs{R}$ and $\attrs{t^i}$, it projects the $A_{i+1}$ attribute value from every tuple $t \in R$ that can join with the $i$-tuple on all the attributes $As$. Let $T_{i+1}$ be a relation containing all $A_{i+1}$ values found. Second, for $R'$, repeat the same, and let $T'_{i+1}$ be a relation containing all $A_{i+1}$ values found. The result of $\extension{\ti{i}}{\A_{i+1}}$ is the intersection of $T_{i+1}$ and $T'_{i+1}$. At line~6-7, for every value, $v$, in $\extension{\ti{i}}{\A_{i+1}}$, it calls \lf recursively with an $(i$+$1)$-tuple, $t^{i+1} = t^i \| v$, by concatenating $t^i$ and $v$. At line~1-2, If $i = |\attrs{\q}|$, the tuple $t^i$ is emitted through the iterator. It is important to note that the main cost of \lf is the cost of the intersections.

\begin{example}
	Fig.~\ref{fig:Running_Example_2_1}(b) shows the steps of \lf on the server $S_0$ with relations as shown in Fig.~\ref{fig:Running_Example_2_1}(a). The input for the initial \lf call is with an empty tuple $t^0$. Assume the order among $\attrs{\q}$ (e.g., $\ord$ is $a \prec b \prec c \prec d \prec e$).

	First, \lf will project the values for the first attribute $a$ by attempting to join with $t^0$. At the server $S_0$, both relations, $R_1$ and $R_2$, have the attribute $a$. Since $t^0$ is empty, it projects $\{1\}$ from $R_1$ and projects $\{1, 4\}$ from $R_2$, the result of the intersection is $\{1\}$ as shown in the relation $T^1$, whose schema is $\{a\}$, in Fig.~\ref{fig:Running_Example_2_1}(b).

	Second, for the tuple $t^1 = (1)$ in $T^1$, it calls \lf in which the 2nd attribute $b$ in order is considered. Note that both relations, $R_1$ and $R_4$, have the attribute $b$. By joining the tuples in $R_1$ with $t^1 = (1)$, it projects the $b$ attribute value, $\{2\}$, with $R_1$ since the corresponding tuple $t \in R_1$ can join with the input tuple $t^1$ on the attribute $a$, and it projects the $b$ attribute values, $\{2\}$, with $R_4$, since it does not have the attribute $a$ to join with $t^1$. The intersection of $b$ attribute values from the two relations is $\{2\}$, as shown in the relation $T^2$ on the schema $(a, b)$ in Fig.~\ref{fig:Running_Example_2_1}(b). The new $t^2$ to be used in the next \lf call becomes $(1, 2)$ on the schema $(a, b)$.

	Fig.~\ref{fig:Running_Example_2_1}(b) shows the results for $T^1$ to $T^5$ by \lf at the server $S_0$. Here, the join result for the hypercube assigned is in $T^5$. It is worth noting that \lf is implemented as a series of iterator to avoid the recursive function call, and every newly generated tuple $t^{i+1} \in T^{i+1}$ is used immediately to generate tuples $t^{i+2}$ without being stored in memory. 
\end{example}

\stitle{HCube Shuffle}~\cite{afrati_optimizing_2011,beame_communication_2013} is one of the state-of-the-art communication methods to evaluate a join query $\q$ in a distributed system by shuffling data in one-round. The main idea is to divide the output of a join query $\q$ into hypercubes with coordinates, and assign one or more hypercubes to one of the $N^*$ servers to process by shuffling the tuples, whose hash values partially matches the coordinate of the given hypercube, to the server. Given a vector $\p = (\p_1, \p_2, \cdots, \p_n)$, where $p_i$ is the number of partitions for the attribute $A_i$ under $\ord$, and $n = |\attrs{\q}|$, hypercubes of $P = \p_1 \times \cdots \times \p_n$ dimension are constructed. It is worth mentioning that $P$ can be larger than $N^*$. Here, a hypercube is identified by an coordinate of $C = (c_1, ..., c_n)$ of $[\p_1] \times \cdots \times [\p_n]$, where $[l]$ represents the range from 0 to $l-1$. Each machine can be assigned one or more hypercubes. \hcube distribute tuples of each relation to machines via shuffling by hashing. For example, let's assume $p = (1, 2, 2, 1, 1)$, which specifies four hypercubes with coordinates $(0, 0, 0, 0, 0)$, $(0, 1, 0, 0, 0)$, $(0, 0, 1, 0, 0)$, $(0, 1, 1, 0, 0)$. The first tuple, (1, 2,1), that appears at the top in the relation $R_1(a, b, c)$, will be shuffling to the servers that are assigned hypercube with coordinate $(0,0,0,\star, \star)$, since $h_{a}(1) = 0$, $h_{b}(2) = 0$, $h_{c}(2) = 0$, where $h_{A_i}$ means the hash function $h_i$ for attribute $A_i$, and $\star$ means any integer.

\begin{example}\label{exp:exp_1} 
	Consider the join query $\q$ (Eq.~(\ref{eq:qexample})) and the 5 relations in Fig.~\ref{fig:Running_Example_1}. Here, $\attrs{\q} = \{a, b, c, d, e\}$. Let $P = N^* = 4$, assume the order among the attributes of $\q$ is $\ord = a \prec b \prec c \prec d \prec e$. Suppose the vector $\p = (\p_1, \p_2, \p_3, \p_4, \p_5) = (1, 2, 2, 1, 1)$ is obtained by the optimizer, where $\p_i$ denotes the number of partitions for the attribute $A_i$. For example, $\p_1$ is for the attribute $a$ because $a$ is the first attribute in $\ord$. The hypercubes based on $\p$ are $[1] \times [2] \times [2] \times [1] \times [1]$. Note that $[l]$ represents a range from 0 to $l-1$. The 4 hypercubes to be assigned to the 4 servers, $S_0$, $S_1$, $S_2$, and $S_3$ are hypercubes with coordinate $C_0 = (0,0,0,0,0)$, $C_1 = (0,0,1,0,0)$, $C_2(0,1,0,0,0)$, and $C_3 = (0,1,1,0,0)$, The tuples in any of the 5 relations will be sent to some hypercubes. Here, suppose that a hash function, $h_i(\cdot)$, is designed for the $i$-th attribute $A$, and the hash function is of $h_i(x) = x \% \p_i$ for this example. The first tuple, (1, 2, 1), that appears at the top in the relation $R_1(a, b, c)$, will be sent to the servers with hypercubes with coordinate $(0,0,0,\star, \star)$, since $h_{a}(1) = 0$, $h_{b}(2) = 0$, $h_{c}(2) = 0$, where $h_{A_i}$ means the hash function $h_i$ for attribute $A_i$, and $\star$ means any integer. The tuples of the 5 relations that are sent to the server $S_0$ are shown in Fig.~\ref{fig:Running_Example_2_1}(a). 
\end{example}	

	\comment{ is shuffled to target server with coordinate $(0,0,0,\star, \star)$, which includes server 0 and server 1. The rest of tuples are shuffled in a similar manner. On server 0, the tuples whose hash of value of attribute $a, b, c, d, e$ correspond to $0, 0, 0, 0, 0$ respectively is received. one of the $P$ servers. Fig~\ref{fig:Running_Example_1} is shuffled to targets servers. i.e., let $h_i(x) = x \% \p_i$,
	
	tuple $(1, 2, 2)$ of relation $\R_1(a, b, c)$, whose hash value is $h_{\pos{a}}(1) = 0, h_{\pos{b}}(2) = 0, h_{\pos{\A}}(2) = 0$, is shuffled to target server with coordinate $(0,0,0,\star, \star)$, which includes server 0 and server 1. The rest of tuples are shuffled in a similar manners. On server 0, the tuples whose hash of value of attribute $a, b, c, d, e$ correspond to $0, 0, 0, 0, 0$ respectively is received.
	
	In Fig.~\ref{fig:Running_Example_2_1}(a), with query $\q$, database $\D$ and \ord specific in Fig~\ref{fig:Running_Example_1}, and a share vector $\p = [1, 2, 2, 1, 1]$ obtained from optimizer. Four servers are partitioned into a hypercube of $[1] \times [2] \times [2] \times [1] \times [1]$. Each tuple of database $\D$ of

\label{exp:hcube} \end{example}
}

After \hcube completes its shuffling by hashing, each server can compute the data assigned to it using an in-memory multi-way join algorithm independently, i.e., \lf, and the union of the results by the servers is the answer for the join query $\q$.
\begin{algorithm}
	[t]

	\KwIn{an $i$-tuple $t^i$, the query $\q$} \KwOut{tuples of $\q$ emitted through iterators}
	
	\If{$i = |\attrs{\q}|$}{ Emit($\ti{i}$); }\Else{

	let $\Rs_{i+1}$ be the set of relations $R$ in $\q$ if $R$ contains the $(i$+$1)$-th attribute $A_{i+1}$ in order; \\
	
	find all $A_{i+1}$ values that can join the input tuple $t^i$, denoted as $\extension{\ti{i}}{\A_{i+1}}$; \\
	\For{\mbox{each attribute value} $v$ \mbox{in} $\extension{\ti{i}}{\A_{i+1}}$}{ {\lf}($\ti{i} \| v$, $\q$); } }
	
	\caption{{\lf}($t^i$, $\q$)}
\label{algo:leapfrog} \end{algorithm}

\stitle{Remark.} Given the two main costs, namely, communication cost (shuffling cost) and computation cost, \hcubej is designed to puts the communication cost at a higher priority and minimizes the communication cost first by optimizing $p$. There is no concern from \hcube on the computation cost of \lf, which does its best to process the query $\q$ over the data shuffled to it.

However, the query $\q$ could be inherently computationally difficult, and the communication cost may not be the dominating factor in distributed join processing as shown in Fig.~\ref{fig:intro_example}(b). A key question we ask is which cost it should minimize. There are several options, (1) the communication cost, (2) the computation cost, and (3) the both. \hcubej takes the first option. However, It is highly likely that the minimization of communication cost leads to high computation cost. In this paper, we study how to optimize query $\q$ by converting it into an equivalent query $\q_i$ with potential higher communication cost and lower computation cost with some additional pre-computing cost such that the total cost is minimal.

\begin{figure}
	[tbp] \center 
	\includegraphics[width=0.95\columnwidth]{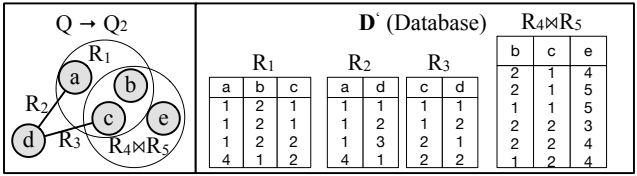} 
	\caption{A query candidate $\Q_i$ which gets the same result of $\q$ in Fig.~\ref{fig:Running_Example_1} by replacing $R_4$ and $R_5$ with $R_4 \bowtie R_5$} \vspace{-0.5cm} 
\label{fig:Running_Example_1_2} \end{figure}

\section{Adaptive Multi-way Join} \label{sec:adj}

In this paper, we study how to minimize the total cost of both communication cost and computation cost together with some additional pre-computing cost. To achieve it, we need a mechanism that allows us to balance the total costs with the condition that the mechanism is cost-effective to achieve the goal of minimization of the total costs.

We discuss our main idea using an example. Consider a join query as $\q = R_1 \bowtie R_2 \bowtie R_3 \bowtie R_4 \bowtie R_5$ (refer to Eq.~(\ref{eq:qexample}) for the details) over the database $D$ shown in Fig.~\ref{fig:Running_Example_1}. Let it be executed by \hcubej, where \hcube shuffles the database $D$, and \lf is deployed on each server to compute the data shuffled to it. Assume, the system finds out that the time spent on \hcube for shuffling tuples is relatively small, while a considerable amount of time is spent on \lf on each server. Furthermore, suppose the system finds out that the computation cost of \lf can be reduced for the same query $\q$ if $R_4 \bowtie R_5$ has already been joined as one relation instead. In other words, let $\q_2 = R_1 \bowtie R_2 \bowtie R_3 \bowtie R_{45}$ where $R_{45} = R_4 \bowtie R_5$, instead of executing $\q$ directly, it is to pre-compute $R_{45}$ first, then execute $\q_2$. Though it would be more expensive to do the pre-computing and shuffle the tuples of $\mc{R}(\q_2)$, which is shown in Fig.~\ref{fig:Running_Example_1_2} ($18$ integers in $R_{45}$, $16$ integers in $R_4$ and $R_5$ in total), it is still preferable to execute the new query $\q_2$ instead of $\q$ to trade the communication cost and pre-computing cost for computation cost, which is the bottleneck. The message by this example is: there is a way to reduce the computation cost at the expense of increased communication cost with some pre-computing cost, and it is possible to minimize the total cost by balancing the computation cost, communication cost, and pre-computing cost.

We give our problem statement below based on the idea presented in the example. Consider a join query $\q = R_1 \bowtie R_2 \bowtie \cdots \bowtie R_m$ (refer to Eq.~(\ref{eq:join})). Let $\Qs$ be a collection of query candidates such as $\Qs = \{\Q_1, \Q_2, \cdots \Q_{|\Qs|}\}$, where $\Q_i = R'_1 \bowtie R'_2 \bowtie \cdots \bowtie R'_l$. Here, $\Q_i$ is equivalent to $\q$ such that $\Q_i$ and $\q$ return same results, $\attrs{\Q_i} = \attrs{\q}$, $l \leq m$, and a relation $R'_j$ in $\Rs(\Q_i)$ is either a relation $R_k$ in $\Rs(\q)$ or a relation by joining some relations in $\Rs(\q)$. Let a query plan be a pair $(\q_i, \ord)$ that consists of a query candidate $\Q_i \in \Qs$, which specifies how to pre-compute relations, and an attribute order $\ord$ for attributes of $\q_i$, which specifies how to join the relations of new query $\Q_i$ using \lf. The problem is to find a query plan such that the total cost for communication, pre-computing, and computation is minimized. This problem is challenging due to the huge search space. For example, there exists  $2^m$ possible combinations of joins to construct a single relation $R'_j$ in total, where $m$ is the number of relations in $\q$, and $n !$ possibilities to order the attributes of $\Q_i$.

In this paper, we propose a prototype system (\apj) that explores cost-effective query plans from a reduced search space. The workflow of our system (\apj) is as follows. First, we shrink the search space according to an optimal hypertree $\T$ constructed for query $\q$ such that search space of candidate relations and attribute order $\ord$ are reduced based on $\T$. Then, we explore cost-effective query plans derived from the $\T$ by considering the cost-effectiveness of trading the computation with communication and pre-computing of each pre-computed candidate relations with the cost model. The cardinality estimation is done via a distributed sampler. Given an optimal query plan $(\Q_i, \ord)$, first, for each relation $R'_j \in \Q_i$ that needs to be joined, we pre-compute and store it. After every $R'_j$ is computed, we execute $\Q_i = R'_1 \bowtie R'_2 \bowtie \cdots \bowtie R'_l$. As shown in Fig.~\ref{fig:intro_example}(b), our approach can significantly reduce the total cost. 

Next, in Sec~\ref{sec:candidate_selection}, we explain how to reduce the search space. Then in Sec~\ref{sec:cost_model} we show how to explore cost-effective query plans based on hypertree $\T$. How to estimate the cardinality via distributed sampling is shown in Sec~\ref{sec:p_estimation}.

\subsection{The Reduced Search Space} \label{sec:candidate_selection}

\begin{figure}
	[t] \centering 
	\includegraphics[width=0.7
	\columnwidth]{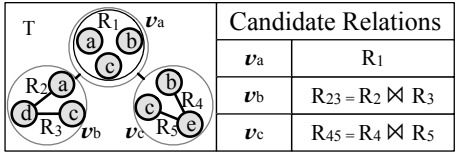} 
	\caption{Hypertree $\T$ and candidate relations \label{fig:Running_example_3}} 
	\vspace{-0.6cm}
\end{figure}

To reduce the search space for selecting an optimal query plan from the collection of query candidates $\Q_i \in \Qs$ and possible attribute orders, we only consider a limited number of joins such that a join (e.g., $R_4 \bowtie R_5$) is as small as possible and could lower join cost of $\Q$. More specifically, we find query candidates that are almost acyclic queries and can be easily transformed from $\Q$. Our intuition is that the computation cost of evaluating an acyclic query is usually significantly smaller than that of evaluating an equivalent cyclic query. Thus an almost acyclic query $\Q_i$ could be easier to evaluate than $Q$.

This is done as follows. First, we represent a given join query $\q$ using its hypergraph representation, $H = (V, E)$. Second, for the hypergraph $H$, we find a hypertree representation, ${\cal T} = (V, E)$, where $V({\cal T})$ is a set of hypernodes and $E({\cal T})$ is a set of hyperedges. Recall that, in the hypergraph $H$, a hypernode represents an attribute, and a hyperedge represents a relation schema. The corresponding hypertree ${\cal T}$ represents the same information. (1) A hypernode in $V({\cal T})$ represents a subset of hyperedges (e.g., relation schemas) in $E(H)$, and it also corresponds to a potential pre-computed relation, which can be computed by joining the corresponding relations of the relation schemas it contains. (2) Hyperedges $E({\cal T})$ of ${\cal T}$ is constructed such that the hypernodes in ${\cal T}$ that contains a common attribute $A$, must be connected in the hypertree ${\cal T}$. 

There are many possible hypertrees for a given hypergraph, we use the one whose maximal size of the pre-computed relation of each hypernode is minimal. This requirement ensures that for any subset of hypernodes $V'(\mc{T}) \subseteq V(\mc{T})$ to be pre-computed, the resulting relations do not incur too much pre-computing and communication overhead in later join query $\q_i$. We find such a hypertree ${\cal T}$ using \ghd (Generalized HyperTree Decomposition)~\cite{gottlob_hypertree_2002}. To bound the maximum size of the pre-computed relation of each hypernode in the worst-case sense, in theory, we can select the one with minimal \kw{fhw} (fractional hypertree width)~\cite{gottlob_hypertree_2016}. Such a hypertree ${\cal T}$ found by \ghd satisfies that $\max_{v \in V({\cal T})} \abs{R_{max}}^{\kw{fhw}}$ is the lowest among all hypertrees, where $\abs{R_{max}} = \max_{R \in \mc{R}(\q)} \abs{R}$. In other words, the size of every pre-computed relation of each hypernode is upper bounded by $\abs{R_{max}}^{\kw{fhw}}$ for the chosen ${\cal T}$ and it is the lowest one among all possible ${\cal T}$. 

\begin{example}
	Consider the join query $\q = \R_{1}(a, b, c) \bowtie \R_{2}(a, d)$ $ \bowtie \R_{3} (c, d) \bowtie \R_{4}(b, e) \bowtie \R_{5}(c, e)$ (Eq.~(\ref{eq:qexample})). Its hypergraph is shown in Fig.~\ref{fig:Running_Example_1}, and its hypertree ${\cal T}$ is shown the leftmost in Fig.~\ref{fig:Running_example_3}. For the hypertree ${\cal T}$, its hypernodes are ${v_a, v_b, v_c}$, where $v_a$, $v_b$, and $v_c$, represent $R_1(a, b, c)$, $R_2(a, d) \bowtie R_3(c, d)$, and $R_4(b, e) \bowtie R_5(c, e)$, respectively. The hyperedges $\{(v_a, v_b), (v_b, v_c)\}$ ensure 1) $T$ is a hypertree 2) For any attribute $A \in \{a, b, c, d, e\}$, e.g., $a$, the hypernodes that contains it, e.g., $v_a, v_b$, are connected. 
\label{exam:htree} \end{example}

As shown in Example~\ref{exam:htree}, the hypertree ${\cal T}$ found from the hypergraph representation for a given join query, $\q = R_1 \bowtie R_2 \bowtie \cdots \bowtie R_m$, has two implications regarding the reduced search space to find the optimal $\Q_i = R'_1 \bowtie R'_2 \bowtie \cdots \bowtie R'_l$, namely, the number of joins and the attribute order.

\stitle{Reducing Numbers of Candidate Relations.} Instead of finding any possible joins to replace a single relation $R'_j$ in $\Q_i$, we only consider the joins represented as hypernodes in the hypertree ${\cal T}$. By pre-computing such joins, query $\Q_i$ is almost acyclic. Consider the hypertree, ${\cal T}$, as shown the leftmost in Fig.~\ref{fig:Running_example_3} for $\q = R_1 \bowtie R_2 \bowtie R_3 \bowtie R_4 \bowtie R_5$. The hypertree ${\cal T}$ has three hypernodes that represent $R_1(a, b, c)$, $R_2(a, d) \bowtie R_3(c, d)$, and $R_4(b, e) \bowtie R_5(c, e)$, respectively. Here, $R_1(a, b, c)$ is a relation appearing in $\q$, and there is no need to join. For the other two hypernodes, there are only 4 choices, namely, not to pre-compute joins, to pre-compute the join of $R_2(a, d) \bowtie R_3(c, d)$, to pre-compute the join of $R_4(b, e) \bowtie R_5(c, e)$, to pre-compute both joins. In other words, by the hypertree, ${\cal T}$, for this example, we only need to consider 4 possible query candidates, which decides whether $R_{23}$ and $R_{45}$ should be pre-computed. The search space of query candidates is significantly reduced to $2^{\abs{V({\cal T})}}$.

\comment{ $\Q_1 = \R_{1}(a, b, c) \bowtie \R_{2}(a, d) \bowtie \R_{3} (c, d) \bowtie \R_{4}(b, e) \bowtie \R_{5}(c, e)$,

$\Q_2 = \R_{1}(a, b, c) \bowtie \R_{23}(a, c, d) \bowtie \R_{4}(b, e) \bowtie \R_{5}(c, e)$ for $\R_{23}(a, c, d) = \R_{2}(a, d) \bowtie \R_{3} (c, d)$,

$\Q_3 = \R_{1}(a, b, c) \bowtie \R_{2}(a, d) \bowtie \R_{3} (c, d) \bowtie \R_{45}(b, c, e)$ for $\R_{45}(b, c, e) = \R_{4}(b, e) \bowtie \R_{5}(c, e)$, and

$\Q_4 = \R_{1}(a, b, c) \bowtie \R_{23}(a, c, d) \bowtie \R_{45}(b, c, e)$ for $\R_{45}(b, c, e) = \R_{4}(b, e) \bowtie \R_{5}(c, e)$. }

\stitle{Reducing Choice of Attribute Orders.} \lf needs to determine the optimal attribute order to expand from $i$-tuple to $(i$+$1)$-tuple. For a query $\q$ with $n$ attributes for $n = |\attrs{\q}|$, there are $n!$ possible attribute orders to consider for any query $\Q_i$ in $\Qs$, which incurs high selection cost. With the hypertree ${\cal T}$, it can reduce the search space to determine an attribute order following a traversal order ($\prec$) of the hypernodes of the hypertree, ${\cal T}$. Consider any hypernodes, $u$ and $v$, in ${\cal T}$, where $u$ appears before $v$ (e.g, $u \prec v$) by the traversal order. First, an attribute that appears in $u$ will appear before any attribute in $v$ that does not appear in $u$. Second, the attributes in a hypernode $v$ can vary if they do not appear in $u$, and can be determined via \cite{chu_theory_2015}.For hypertree ${\cal T}$ shown in the leftmost of Fig.~\ref{fig:Running_example_3}, let's assume the traversal order among the hypernodes are $v_a \prec v_b \prec v_c$. A valid attribute order is $a \prec b \prec c \prec d \prec e$, and an invalid attribute order is $a \prec b \prec e \prec d \prec c$. The rationale behind such reduction is that the attributes inside a hypernode are tightly constraint by each other, while attributes between two hypernodes are loosely constraint, thus when following a traversal order, the attributes of $A_1, ..., A_{n-1}$ are more likely to be tightly constraint, which results in less intermediate tuples $t^1, ..., t^{n-1}$ of $T^1, ..., T^{n-1}$ respectively during \lf. An experimental study in Sec.~\ref{sec:Experiment} confirms such intuition. By adopting such order, the search space of attribute order is reduced from $O(n!)$ to $O(\abs{V({\cal T})}!)$, where $\abs{V({\cal T})} < n$.
%

\subsection{Finding The Plan \label{sec:cost_model}} 

In this section, we discuss how to find a good plan from the reduced search space.

\stitle{The Optimizer.} Let $n^* = \abs{V({\cal T})}$, a naive approach finds the optimal plan by considering every combination of query candidates that form from candidate relations and every traversal orders, which are $O(2^{n^*} \times n^*!)$ plans in total. It is worth mentioning that calculating the cost for each plan could be costly as well. Thus finding plans by such a naive approach is not feasible.

We propose an approach to find good plans by exploring effective candidate relations in terms of trading the computation with communication. Recall that, pre-computing candidate relations could reduce the computation cost but increase the communication cost, and bring additional pre-computing cost. By finding the candidate relations that have a large positive utility in terms of reducing computation cost, we can effectively trade the computation cost with communication cost. 

\begin{figure}
	[tbp] 
	\begin{center}
		\begin{tabular}
			[t]{c} \hspace{-0.5cm} \subfigure[$Q_5$] { 
			\includegraphics[width=0.49\columnwidth]{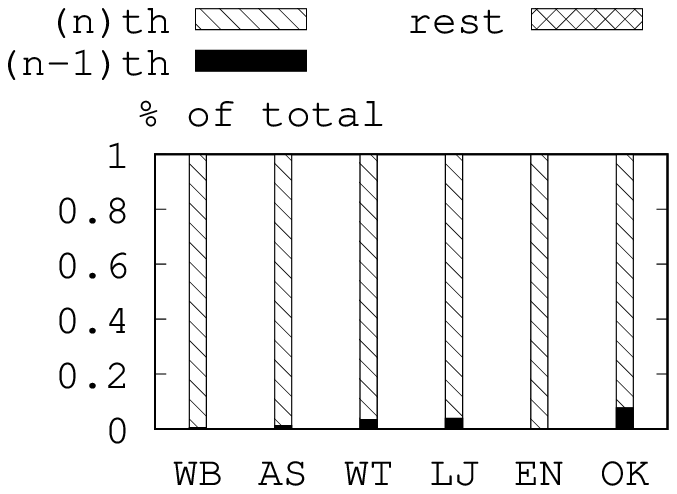} } \subfigure[$Q_6$] { 
			\includegraphics[width=0.49\columnwidth]{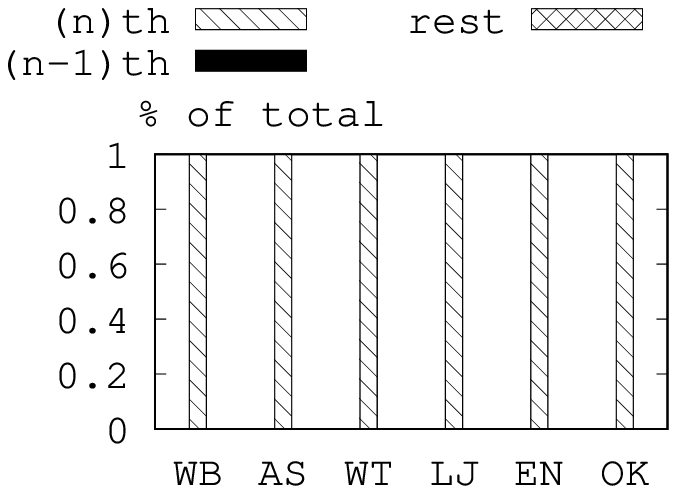} } 
		\end{tabular}
	\end{center}
	
	\vspace{-0.4cm} 
	\caption{Percentages of intermediate tuples to extends during traversing $n-th$ node, $(n-1)-th$ node, and the rest of the node using two join queries, $Q_5$ and $Q_6$ (same as Fig.~\ref{fig:intro_example}).\label{fig:last_node}}\vspace{-0.4cm} 
\end{figure}

Let $C$ be the set of candidate relations to pre-compute, $O$ be the traversal orders, $cost_M(C)$, $cost_C(C)$, and $cost_E^i(C,O)$ be the cost of pre-computing cost, communication cost, and the computation cost of steps that extends to attributes of $i$-th traversed nodes in \lf. It is worth noting that in complex join, the last few steps of \lf usually dominate the entire computation cost due to a large number of partial bindings to extend~\cite{chu_theory_2015}, and reducing such cost by pre-computing $R_v$ usually has maximum benefits in terms of reducing computation cost. An example is also shown in Fig.~\ref{fig:last_node}. Assuming we have an empty $C$ and empty $O$. For each candidate relations $R_v$, where $v \in V({\cal T})$, we try to explore its maximum utility by setting last traversed node of $O$ to $v$. Then we compare the cost of pre-computing $R_v$ and not pre-computing $R_v$, which are $cost_M(R_v) + cost_C(C \cup R_v)+ cost_E^{n*}(C \cup R_v, O)$ and $cost_C(C)+ cost_E^{n*}(C, O)$ respectively, with the cost of current optimal candidate relation $R_{v*}$ in terms of cost. We only consider computation cost last steps of $\lf$, as it usually dominates the entire computation cost. After that, we can proceed to the next round of selecting $R_u$ from the remaining candidate relations in a similar fashion and determining which node $u$ the $(n-1)$-th traversed node and whether $R_u$ should be pre-computed.

The detailed procedure is described in Alg.~\ref{algo:optimizer}. Here, in lines 3-14, we gradually determine all candidate relations and the traversal order in reverse order. In lines 5-13, we find the next candidate relations. The if condition in line 6 is used to ensure that only $O$ that could be extended to valid traversal order, which is described in the last section, is considered. In lines 7-13, we compare the cost of pre-computing $R_v$ and not pre-computing $R_v$ with the cost of current optimal candidate relation $R_{v^*}$. Notice that, in i-th iteration, we only need to compute the $cost_E^i(C',O')$, as the computation cost of $cost_E^{i'}(C',O')$ is the same for all candidates relations for $i' > i$.
\begin{lemma}
	Cost of Alg.~\ref{algo:optimizer} is $O(\frac{1}{2} (2n^*)(2n^* -1)L)$, and $L$ is a large constant factor that is related to the cost of estimating the $cost_M$, $cost_C$, and $cost_E$.
\end{lemma}

\begin{algorithm}[tbp]
\DontPrintSemicolon
\KwIn{Query $Q$}
\KwOut{The optimal query plan $(\Q_i, \ord)$}
find optimal hypertree ${\cal T}$ for $\q$\\
let $C = \emptyset$, $O = \emptyset$, $V = V({\cal T})$ \\

\While{$V \neq \emptyset$}{
$C^* = C$, $O^* = O$, $cost = \inf$, $v^* = null$, $i=n^*$\\
\For{$v \in V$}{
		\If{\text{any two nodes in $V \setminus v$ are connected}}{
		$O' = O.add(v), C' = C \cup R_v$\\
		$cost' = cost_C(C) + cost_E^i(C, O')$\\
		$cost'' = cost_M(R_v) + cost_C(C') + cost_E^i(C', O')$\\
		\If{$cost' < cost$}{
			$C^* = C$, $O^* = O'$, $cost = cost'$, $v^* = v$\\
			}\ElseIf{$cost'' < cost$}{
			$C^* = C'$, $O^* = O'$, $cost = cost''$, $v^* = v$\\
			}
		}
	}
	$i=i-1$, $V.remove(v^*)$, $C = C^*$, $O = O^*$\\
}
convert $C,O.reverse()$ to $\Q_i, \ord$\\
\Return $(\Q_i, \ord)$;

\caption{Optimizer($\q$, $D$)}
\label{algo:optimizer}
\end{algorithm}

\stitle{Computing the Cost.} Next, we discuss how to compute pre-computing cost $cost_M$, communication cost $cost_C$, and computation cost of $i-th$ step in \lf $cost_E$. We focus on computing $cost_C$ and $cost_E$, as $cost_M$ is just a combination of $cost_M$ and $cost_E$.

$cost_C(C)$ measures the communication cost of shuffling relations of $R_v \in C$ and remaining relations of $u \in V({\cal T})$ that are not pre-computed $R_u$ in terms of seconds needed to transmit them across servers. Let us denote such collection of relations by ${\cal R}_C$. Recall that \hcube has a parameter $p$, which determines the numbers of partitions on attribute $A \in \attrs{Q}$ and is related to how tuples are shuffled to each servers. Given a $p$, for each relation $R \in {\cal R}_C$, each tuple $t \in R$ will be sent to $dup(R, p) = \prod_{A \in \attrs{\q} \setminus \attrs{R}} p_A$ servers following the rules of \hcube, where $p_A$ denotes numbers of partitions on attribute $A$. And, we can represent $cost_C(C)$ as $\frac{\sum_{R \in {\cal R}_C} \abs{R} \times dup(R, p)}{\alpha}$, where $\alpha$ is the number of tuples transmitted per seconds. Here, $p$ is a parameter of \hcube and it needs to be optimized to minimize $cost_C(C)$ under the constraints 1) numbers of partition for each attribute should $\geq$ 1; 2) on average, the total amount of data a server received should be less than memory size $M$ of the server, which translates to $M - \sum_{R \in \Q_i} \abs{R} \times frac(R, p) \geq \boldsymbol{0}$. Here, $frac(R, p)$ denotes the average percentage of $R$ will be sent to a server, which is $\frac{1}{\prod_{A \in \attrs{R}} p_A}$. The optimization program is as follows: 
\begin{equation}
	\setlength\arraycolsep{1.5pt} 
	\begin{array}{l@{\quad}
		r c l c l} \mathrm{minimize} & cost_C(C) & & & & \\
		\mathrm{s.t.} & p & - & \textbf{1} & \geq & \textbf{0} \\
		& M & - & \sum_{R \in \Q_i} size(R) \times frac(R, p) & \geq & 0 
	\end{array}
\end{equation}

\noindent By solving above optimization program, we can obtain $cost_C(C) = \sum_{R \in {\cal R}_C} \abs{R} \times dup(R, p)$.

$cost_E^i(C,O)$ measures the computation cost of steps that extends attributes of $i$-th traversed nodes in \lf. Recall that \lf gradually extends i-tuple $t^i \in T^i$ to $(i+1)$-th tuples $t^{i+1}$, $(i+2)$-th tuples, ..., $n$-th tuples. As single node $v \in V({\cal T})$ might contains several attributes, and extending one node $v$ might corresponds to extending several attributes, for simplicity, we use $T^{v_i}$ to denote the tuples of partial binding of attributes are from $v_1$, $v_2$, ..., $v_{i}$, where $v_i$ is the i-th traversed node. Thus, we can represent the cost of extending attributes of $i$-th traversed nodes in \lf, $cost_E^i(C,O)$,  as $\frac{\abs{T^{v_{i-1}}}}{\beta^i \times N^*}$, where $\abs{T^{v_{i-1}}}$ is the numbers of partial bindings whose attributes are from $v_1, ..., v_{i-1}$, $\beta^i$ is numbers of partial bindings extended per seconds per server, and $N^*$ is the number of servers. Notice that $\beta^i$ can be significantly higher if $v_i$ is pre-computed.

$cost_M$ measures the pre-computing cost of $R_v$. Let $\lambda(v)$ be the relations of a node $v$ in $V({\cal T})$, $cost_M$ consists of the communication cost of shuffling $\lambda(v)$ and computation cost of $\bowtie \lambda(v)$, which can be computed using above methods for computing $cost_C(C)$ and $cost_E^i(C,O)$.

In the above calculation, $\alpha$ can be regarded as a constant that measures the communication performance of the cluster. More specifically, we can measure it by randomly generating tuples of size $k$, which is to be shuffled to random servers in the cluster, and recording the time $t$ to shuffling $k$ tuples to their destination, where $\alpha = \frac{k}{t}$. $\beta$ can be estimated by sampling some partial bindings, extending them, and taking the average of their extending time. More specifically, if $v_i$ is pre-computed, the main cost of extending a partial binding is querying the trie for candidate values, thus $\beta^i$ is a constant that can be pre-measured as $\frac{k}{t}$ by recording the time $t$ to perform $k$ query on a trie of size $\abs{R_{v_i}}$. It is worth noting that we can pre-measure $\beta^i$ on trie of various sizes. If $v_i$ is not pre-computed, we set $\beta^i$ by reusing statistics gathered during sampling, which is to be explained in the next section. More specifically, let the total numbers of extension performed during sampling be $k$ and aggregated extension time be $t$, we set $\beta^i = \frac{k}{t}$.

\section{Estimating Cardinality Via Distributed Sampling} \label{sec:p_estimation} 

In this section, we discuss how we perform cardinality estimation via distributed sampling and why we choose sampling-based approaches to estimate cardinality.

\stitle{Why Sampling.} An accurate cardinality estimation is crucial for the optimizer to choose a good query plan \cite{leis2015good}. Currently, there are two styles to do cardinality estimation: 1) sketch-based approaches 2) sampling-based approaches.

Theoretical \cite{ioannidis1991propagation} as well as empirical \cite{leis2015good} work has shown that existing sketches-based approaches, which utilize fixed-size, per-attribute summary statistics (histograms) with strong assumptions (uniformity, independence, inclusion, ad hoc constants) to estimate cardinalities, often return estimations with large errors, especially on complex joins with more than 2 relations. Such error has been shown to lead to sub-optimal plans that are up to $10^2$ slower than optimal plans in empirical study work \cite{leis2015good}. For sampling-based approaches, promising result is shown in \cite{leis2017cardinality} that sampling-based approaches could produce estimations that are orders of magnitude more accurate than sketch-based approaches in a reasonable time by performing a sequence of index join with samples. In summary, sketch-based approaches often incur less overhead than sampling-based approaches when performing estimations, but sampling-based approaches usually return estimations with much fewer errors.

As our work targets complex join, which usually is long-running tasks and the additional cost brought by sampling is negligible compared to its benefits in reducing queries' running time, we choose to estimate cardinality via sampling.

\stitle{Estimating Cardinality Via Sampling.} Given a query Q, whose result is $T$, we want to estimate $\abs{T}$. Let $T_{A = a}$ be result tuples in $T$ whose value on attribute $A$ is $a$, we can express $T$ as follows.
\begin{equation}
	\abs{T} = \sum_{a \in val(A)} \abs{T_{A = a}} = \abs{val(A)} \times \overline{\abs{T_{A = a}}}
\end{equation}

\noindent where $val(A)$ is the collection of values of $A$ in $T$, and $\overline{\abs{T_{A = a}}} = \frac{\sum_{a \in val(A)} \abs{T_{A = a}}}{\abs{val(A)}}$. Suppose $\abs{val(A)}$ is known, then we need to estimate $\overline{\abs{T_{A = a}}}$ to obtain an estimation of $\abs{T}$. To estimate $\overline{\abs{T_{A = a}}}$, let $a$ be a randomly selected value from $val(A)$. Let $X$ be the random variable that is $\abs{T_{A = a}}$, and $\mu$ = $\mathbf{E}[X]$ = $\overline{\abs{T_{A = a}}}$. 

Suppose we wish to estimate $\mu$. We simply choose k independent values $a_1, a_2, ..., a_k$ from $val(A)$ with associated random variables $X_1, X_2, ..., X_k$. Define $\bar{X} = \frac{1}{k} \sum_{i < k} X_i$ as our estimate. The generalized Chernoff-Hoeffding bounds \cite{hoeffding1994probability} give gurantees on $\bar{X}$, as follows.
\begin{lemma}
	Let $X_1, X_2, ..., X_k$ be independent random variables with $X_i \in [0, b]$, where $b$ is the maximum values $X_i$ can take. Define $\bar{X} = \frac{1}{k} \sum_{i < k} X_i$. Let $\mu$ = $\mathbf{E}[X]$. Then for $p \in [0,1]$, we have \[ PR\{\abs{\bar{X} - \mu} \geq pb \} \leq 2exp(-2kp^2) \] Hence, if we set $k = \lceil  -0.5p^{-2}ln(2/\delta) \rceil$, then $PR\{ \abs{\bar{X} - \mu} > pb \} < \delta$. In other words, for k samples, with confidence at least $1-\delta$, the error rate, which measures the deviation of $E[\bar{X}]$ in terms of b is at most $p$.
\end{lemma}

In practice, we can easily obtain $val(A)$ by performing intersections over relations of $Q$ that contains $A$ in their schemas, which is $\bigcap\limits_{R \in Q \wedge A \in \attrs{R}} \Pi_{A} R$. We can obtain $\abs{T_{A=a}}$ for any $a$ chosen from $A$ by performing an \lf starting from $A$ with attribute on $A$ being fixed as $a$, which obtains $T_{A=a}$.

\stitle{Distributed Sampling.} A naive approach parallelize the sampling process described above by utilizing \hcube directly. More specifically, it first shuffling the relations of $Q$ into servers using \hcube such that each server can perform the sampling on its own based on tuples on it, then on each server, the sampling process described in the above paragraph is performed. However, such naive approaches would shuffle many unnecessary tuples during \hcube, as only a small fraction of $val(A)$, and performing \lf for them probably will not involve all tuples of every relation in $Q$.

We can reduce such costs by reducing the database first before all relations in it are shuffled by \hcube. First, we find all relations ${\cal R}$ in a database whose schema contains $A$, and compute a projected relation for each of them $\Pi_A R , R \in {\cal R}$. Then, for all $R \in {\cal R}$, we shuffle their $\Pi_A R$ such that we can compute the intersection of them and obtain $val(A)$. Then, from $val(A)$, we randomly select some samples $S'$. Next, we reduce the original database by performing semi-join between $S'$ and $R \in {\cal R}$ to filter unpromising tuples. Finally, we shuffle the reduced database instead of the original database, and perform sampling on it.

\section{Implementation \label{sec:hcubeopt}} We implemented a prototype system in Spark, which is the de-facto platform to perform large scale analytic tasks.

\stitle{Optimizing HCube.} Previously, \hcube is implemented as a sequence of \textsf{map} and \textsf{reduce} stage \cite{afrati_optimizing_2011}, where \textsf{map} stage marks the destination coordinate for each tuple and \textsf{reduce} stage shuffles each tuple to their corresponding servers. Such implementation suffers from significant performance loss due to overwhelming amount of tuples being shuffled. To reduce the cost of \hcube, the key is to reduce the cost of expensive shuffling. A solution is to pull the tuples in blocks from remote machines directly instead of shuffling tuples one by one, which bypass shuffling process. The new \hcube proceed in two steps: 
\begin{itemize}
	\item Group all tuples from the same relation and with the same hash values under the \hcube's hash function into a block and tagged that block with that has values. 
	\item For each server, it pulls the entire block of each relation whose hash values ``fits" its own coordinate in blocks from remote machines. 
\end{itemize}
 We next use an example to better illustrate the idea.  
\begin{example}\label{exp:push} 
	Let's take query in Fig.~\ref{fig:Running_Example_1} whose share $p = (1, 2, 2, 1, 1)$, which result in four servers with coordinate $(0, 0, 0, 0, 0)$, $(0, 1, 0, 0, 0)$, $(0, 0, 1, 0, 0)$, $(0, 1, 1, 0, 0)$. For the relation $R_3$ with schema $R_3(c, d)$, its tuples will be split into two blocks, where $(1, 1)$, $(1, 2)$ will be in block $B(0,0)$, and $(2, 1)$, $2, 2$ will be in block $B(1,0)$ as their hash value for $c,d$ is $1,0$ and $0,0$ respectively. And, the servers with coordinate $(0, 0, 0, 0, 0)$, $(0, 1, 0, 0, 0)$ will pull block $B(0,0)$ as their coordinate on $c$ and $d$ is $0,0$. Similarly, servers with coordinate $(0, 0, 1, 0, 0)$, $(0, 1, 1, 0, 0)$ will pull block $B(1,0)$. 
\end{example}
 A further benefit that this new \hcube implementation has is that it allows us to do some preprocessing works on a block level. More specifically, we can reduce the cost of constructing the trie of local database in each machine by pre-build the trie for each block of every relation.

\section{Related work} \label{sec:relate} 
Our work is related to previous works from three areas: multi-way join on a single machine, distributed multi-way join, and cardinality estimation.

\stitle{Multi-Way Join on a Single Machine.} Optimizing the computation cost of a multi-way join has been studied for decades. Traditional multi-way join~\cite{selinger_access_1979} is based on relational algebra (RA) --- an RA expression of a multi-way join represents a sequence of binary joins, i.e., sort-merge join. The recently emerged AGM bound~\cite{atserias_size_2008, grohe_constraint_2014} on the worst-case output size of a multi-way join provides a standard to evaluate the computation efficiency of a join algorithm. In the worst-case, using traditional binary joins is suboptimal while worst-case optimal join algorithms such as NPRR~\cite{ngo_worst-case_2012}, Generic Join~\cite{ngo2014skew}, Leapfrog\cite{veldhuizen_triejoin:_2014} are optimal. To improve the efficiency of worst-case optimal join algorithm for general case rather than worst-case, EmptyHeaded \cite{aberger_emptyheaded:_2017} combines binary join and worst-case optimal join via tree decomposition~\cite{gottlob_hypertree_2002, gottlob_hypertree_2016}, and yannakakis algorithm \cite{yannakakis_algorithms_1981}, which improves the computation efficiency at a great cost of memory consumption. To overcome the memory issue of the EmtpyHeaded, CacheTrieJoin~\cite{kalinsky2016flexible} is proposed, which incorporates multi-level cache into \lf. However, it is difficult to set the size of the cache for each level and the total amount of the cache.

\stitle{Distributed Multi-Way Join.} Traditional multi-way join in the distributed platform such as Spark \cite{armbrust_spark_2015}, consists of a sequence of distributed binary joins, such as distributed sort-merge join. They suffer from high communication cost for shuffling intermediate results when processing complex join queries. Such heavy communication cost can be reduced by one round multi-way join method \hcube~\cite{afrati_optimizing_2011, beame_communication_2013}, which avoid shuffling of intermediate results. The combination of \hcube and \lf forms the \hcubej \cite{chu_theory_2015}, which processes the complex join queries effectively. However, when communication cost has been well optimized, the computation cost becomes the new bottleneck. Also, simply combining \hcube and optimized version of \lf, such as CachedTrieJoin, helps little, as it prioritizes the memory usage for \hcube over memory usage for CacheTrieJoin. Compared to previous work, we trying to co-optimize pre-computing, communication, and computation cost via introducing effective partial results.

\stitle{Cardinality Estimation.} The estimation of cardinality methods can be roughly classified into two classes: 1) sketches based, which use statistics of the database to estimate the cardinality of the query, see \cite{leis2015good} as an entry, 2) sampling-based, which estimates the cardinality by sampling over the database according to query, see \cite{chen_two-level_2017,leis2017cardinality} as an entry. It has been shown that the estimation of the sketch-based method could be orders of magnitude deviate from the ground truth \cite{leis2017cardinality, leis2015good} on complex join.



\section{Experiments} \label{sec:Experiment}

\subsection{Setup} \label{sec:Exp_Setup}

\begin{figure}[tbp]
    \centering
    \includegraphics[width=0.8\columnwidth]{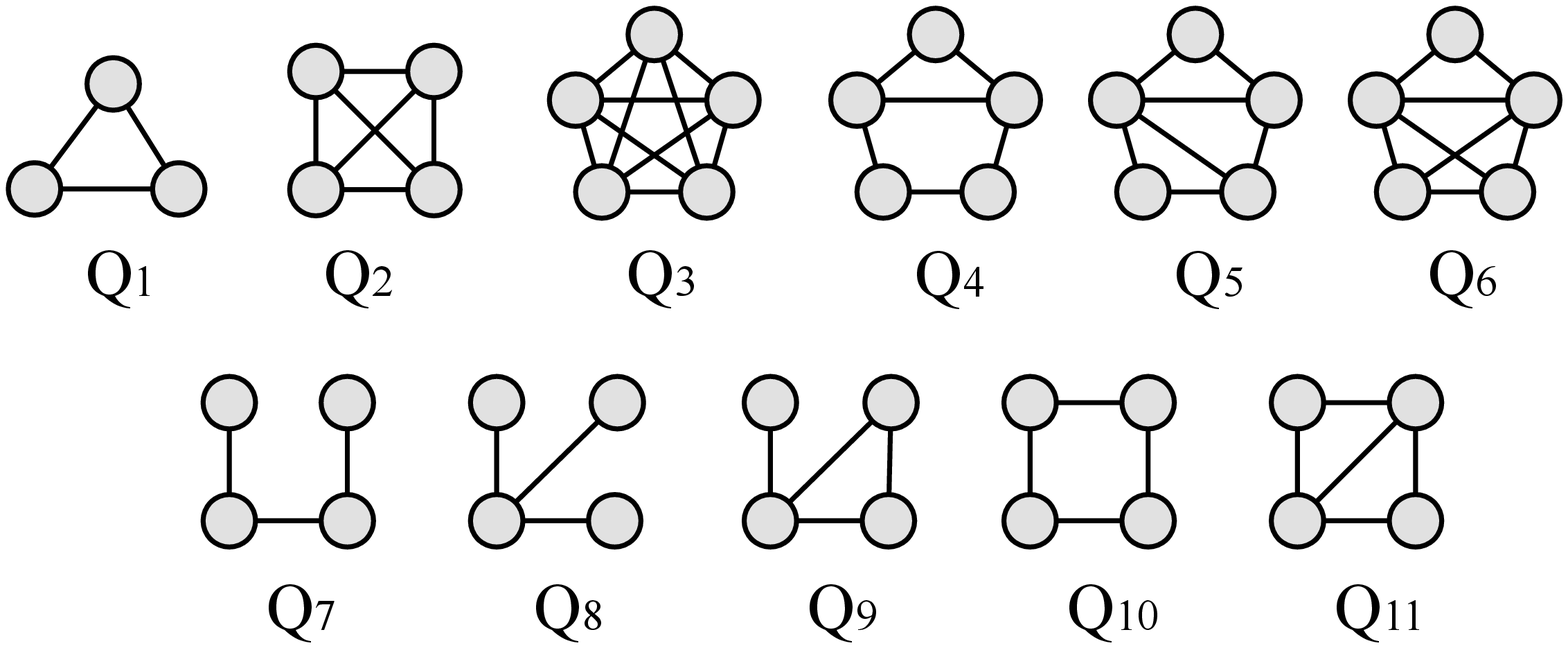}
    \caption{Queries~\label{fig:Pattern}}
    \vspace{-0.1cm}
\end{figure}

\stitle{Queries.} We study complex join queries used in the previous work \cite{chu_theory_2015, kalinsky2016flexible, ammar_distributed_2018}. The queries used are for subgraph queries with nodes in the range of 3-5 nodes.  The queries studied are  shown in Fig.~\ref{fig:Pattern}.  We report the experimental studies for the representative queries from $Q_1$ to $Q_6$, which are not easy to compute.  We omit the results for $Q_7$ to $Q_{11}$, as they can be computed fast, and  the performance of these queries are very similar among the approaches being tested. 
\begin{equation}
	\small
	\nonumber 
	\begin{split}
		\q_1~{}\mbox{:-}~{}& \R_{1}(a, b) \bowtie \R_{2}(b, c) \bowtie \R_{3} (a, c) \\
		\q_2~{}\mbox{:-}~{}& \R_{1}(a, b) \bowtie \R_{2}(b, c) \bowtie \R_{3} (c, d) \bowtie \R_{4}(d, a) \bowtie \R_{5}(a, c) \\&
		\bowtie \R_{6}(b,d) \\
		\q_3~{}\mbox{:-}~{}& \R_{1}(a, b) \bowtie \R_{2}(b, c) \bowtie \R_{3} (c, d) \bowtie \R_{4}(d, e) \bowtie \R_{5}(e, a) \\& 
		\bowtie \R_{6}(b,d) \bowtie \R_{7}(b,e)\bowtie \R_{8}(c,a)\bowtie \R_{9}(c,e) \\&
		\bowtie \R_{10}(a,d) \\
		\q_4~{}\mbox{:-}~{}& \R_{1}(a, b) \bowtie \R_{2}(b, c) \bowtie \R_{3} (c, d) \bowtie \R_{4}(d, e) \bowtie \R_{5}(e, a) \\&
		\bowtie \R_{6}(b, e)\\
		\q_5~{}\mbox{:-}~{}& \R_{1}(a, b) \bowtie \R_{2}(b, c) \bowtie \R_{3} (c, d) \bowtie \R_{4}(d, e) \bowtie \R_{5}(e, a) \\& 
		\bowtie \R_{6}(b, e) \bowtie \R_{7}(b, d)\\
		\q_6~{}\mbox{:-}~{}& \R_{1}(a, b) \bowtie \R_{2}(b, c) \bowtie \R_{3} (c, d) \bowtie \R_{4}(d, e) \bowtie \R_{5}(e, a) \\& 
		\bowtie \R_{6}(b, e) \bowtie \R_{7}(b, d) \bowtie \R_{8}(c, e) \\
	\end{split}
\end{equation}

\begin{table}
	[tbp] 
	\begin{tabular}
		{ccccccc} \toprule Dataset & \WB & \AS & \WT & \LJ & \EN & \OK \\
		\midrule $\abs{R}$ ($\times 10^6$) & 13.2 & 22.1 & 50.9 & 69.4 & 183.9 & 234.4 \\
		Size (MB) & 101.5 & 169.3 & 388.2 & 529.2 & 1370.0 & 1788.1 \\
		\bottomrule
	\end{tabular}
	\caption{Datasets.\label{tab:dataset} } \vspace{-0.4cm} 
\end{table}

\stitle{Datasets.} Following \cite{chu_theory_2015, kalinsky2016flexible}, we construct the database using the real large graph, where each graph is regarded as a relation with two attributes. The statistic of the graphs is shown in Table~\ref{tab:dataset}. For each ``test-case" that consists of a database and a query, the database is constructed by allocating each relation of the query with a copy of the graph. We select $6$ commonly used graphs from various domains. \WB(web-BerkStan) is a web graph of Berkeley and Stanford. \AS(as-Skitter) is an internet topology graph, from traceroutes run daily in 2005. \WT(wiki-Talk) is a Wikipedia talk (communication) network. \LJ(com-LiveJournal) is a LiveJournal online social network. \EN(en-wiki2013) represents a snapshot of the English part of Wikipedia as of late February 2013. \OK(com-Orkut) is an Orkut online social network. Their statistical information is listed in Table~\ref{tab:dataset}. EN can be downloaded from the link \footnote{http://law.di.unimi.it/webdata/enwiki-2013/}, while the rest of the graphs can be downloaded from SNAP\footnote{\url{https://snap.stanford.edu/data/index.html}}.

 \stitle{Competing Methods.} We compare \apj with four state-of-the-art multi-way join methods in the distributed environment. 
\begin{itemize}
	\item \kw{SparkSQL}\cite{armbrust_spark_2015}: The state-of-the-art multi-round multi-way join framework on Spark, which performs multi-way join based on decomposing the query into smaller join queries, and combining intermediate relations in a pairwise way.
	\item \hcubej~\cite{chu_theory_2015}: The state-of-the-art one-round multi-way join framework that utilizes a one-round shuffling method \hcube and the worst-case optimal join \lf
	\item \kw{HCubeJ+Cache}\cite{kalinsky2016flexible}: The state-of-the-art one-round multi-way join framework that utilizes a one-round shuffling method \hcube and adopt an optimized \lf with cache\cite{kalinsky2016flexible}.
	\item \kw{BigJoin}~\cite{ammar_distributed_2018}: The state-of-the-art multi-round distributed multi-round multi-way join framework, which parallelizes \lf. 
\end{itemize}

\stitle{Evaluation Metrics.} We used wall clock time to measure the cost of an algorithm with the time of starting up the system and loading the database into memory excluded. If an approach failed in a test-case due to insufficient memory, the figure will show a space instead of a bar in the corresponding location of the figure. If an approach failed in completing the test-case within 12 hours, we show a bar reaching the frame-top.

\stitle{Parameter Setting.} We set $\alpha$ of \apj by pre-measuring the communication performance of the cluster based on Sec.~\ref{sec:cost_model}. We set the numbers of samples to be $10^5$ as it achieves a balance between accuracy and cost based on our experiments. We set $\beta$ based on Sec.~\ref{sec:cost_model} for each test-case by reusing statistics during sampling of each test-case. For competing methods, we use their default settings.

\stitle{Distributed Settings.} All experiments are conducted on a cluster of a master server and $7$ slave servers ($2$ $\times$ Intel Xeon E5-2680 v4, $176$ gigabytes of memory, interconnected via 10 gigabytes Ethernet). All methods are deployed on Spark 2.2.0. For Spark, we create $28$ workers from $7$ slave servers, where each worker is assigned $7$ cores and $28$ gigabytes of memory. Each core of the worker can be assigned a hypercube in \hcube.

\subsection{The Performance of ADJ} \label{sec:Performance_ADJ} In this section, we investigate the performance of \apj.

\begin{figure}[tbp]
    \centering
     \hspace*{-0.5cm}
    \includegraphics[width=0.85\columnwidth]{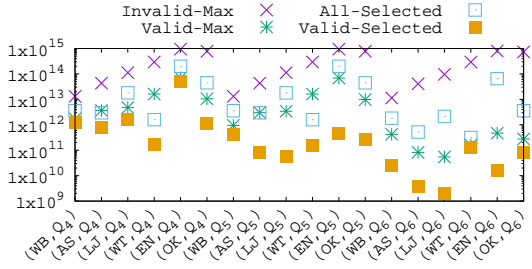}
    \vspace{0.1cm}
    \caption{Effectiveness of attribute order pruning.~\label{fig:valid_invalid_order}}
    \vspace{-0.5cm}
\end{figure}

\stitle{Effectiveness of Attribute Order Pruning.} In this test, we compare the number of intermediate tuples generated during \lf under valid attribute order and invalid attribute order on test-cases using $Q_4-Q_6$ over all datasets. We omit $Q_1-Q_3$, as their intermediate tuples are constant under any attribute order. The results are shown in Fig.~\ref{fig:valid_invalid_order}, where \textsf{Invalid-Max} denotes the attribute order that results in the maximum number of intermediate tuples among all invalid orders. \textsf{Valid-Max} denotes the attribute order that results in the maximum number of intermediate tuples among any valid attribute orders. \textsf{All-Selected} denotes the attribute order selected by \hcubej \cite{chu_theory_2015}, which select the attribute order from all attribute order. \textsf{Valid-Selected} denotes the attribute orders selected by \apj. It can be seen that in terms of the maximum number of intermediate tuples produced, valid attribute orders perform better than invalid attribute orders across all test-case. Also, we can see that selecting the attribute order from only valid attribute orders can produce a better attribute order than considering all attribute orders. This experiment confirms that the effectiveness of our heuristic in selecting good attribute orders and pruning non-effective attribute orders.

\begin{figure}
	[tbp] 
	\begin{center}
		\begin{tabular}
			[t]{c} \subfigure { 
			\includegraphics[width=0.6\columnwidth]{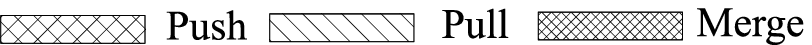} } \addtocounter{subfigure}{-1} 
					\vspace{-0.2cm}
		\end{tabular}
 
		\begin{tabular}
			[t]{c} \hspace{-0.5cm} \subfigure[Communication] { 
			\includegraphics[width=0.53\columnwidth]{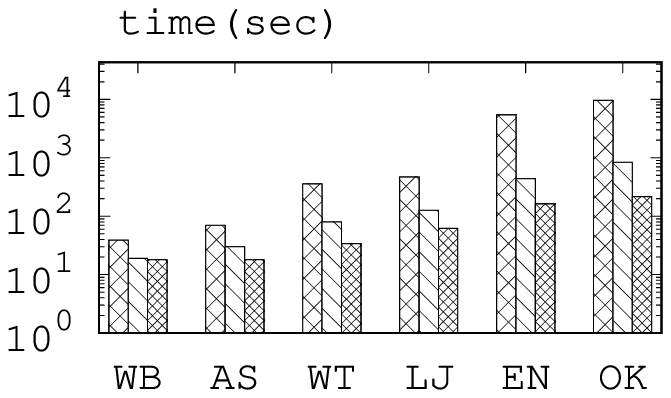} } \hspace{-0.8cm} \subfigure[Computation] { 
			\includegraphics[width=0.53\columnwidth]{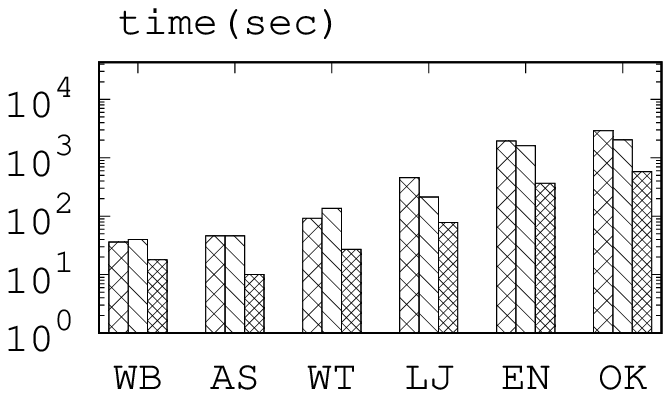} } 
		\end{tabular}
	\end{center}
	\vspace{-0.4cm}

	\caption{Comparison of different implementation of \hcube.  \label{fig:exp_shuffle}.} \vspace{-0.4cm} 
\end{figure}

\stitle{Effectiveness of Optimizations on HCube.} In this test, we compare the effectiveness of the techniques proposed for optimizing the performance of \hcube. We denote the original \hcube implementation by \kw{Push}, our optimized \hcube implementation by \kw{Pull}, and our optimized \hcube implementation with tries pre-constructed by \kw{Merge}. We run test-cases that consist of all datasets and query $Q_2$, and compare the communication cost and cost, where the results are shown in Fig.~\ref{fig:exp_shuffle}. In terms of communication cost, \kw{Pull} and \kw{Merge} outperform \kw{Push} by up to two orders of magnitude. And, \kw{Merge} outperforms \kw{Pull}, as the block that contains one trie, which can be implemented using three arrays, are easier to serialize and deserialize than the block that contains many tuples. In terms of computation cost, \kw{Push} and \kw{Pull} are similar, and \kw{Merge} outperforms the other two methods by up to an order of magnitude as tries has already been pre-constructed before \hcube. This experiment shows that our proposed techniques for \hcube can significantly reduce communication and some computation cost. 
\begin{figure}
	[tbp] 
	\begin{center}
		\begin{tabular}
			[t]{c} \hspace{-0.5cm}
			
			\subfigure[Time] { 
			\includegraphics[width=0.53\columnwidth]{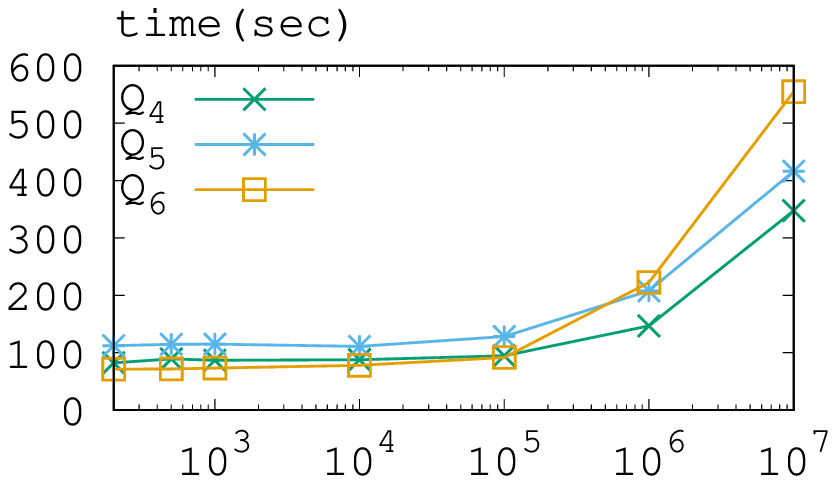} } \hspace{-0.8cm} \subfigure[Max $D$] { 
			\includegraphics[width=0.53\columnwidth]{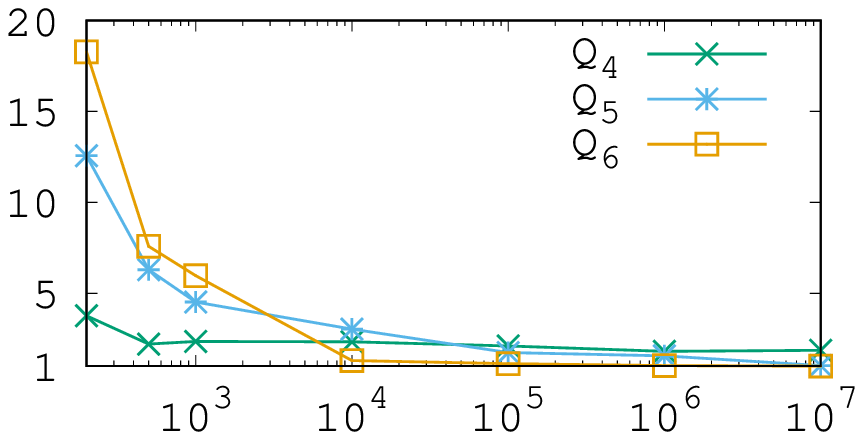} } 
		\end{tabular}
	\end{center}
	
	\vspace{-0.4cm}
	
	\caption{Cost and accuracy of Sampling Process \label{fig:cost_accuracy}} \vspace{-0.6cm} 
\end{figure}

\stitle{Cost and Accuracy of Sampling Process.} In this test, we show that a relatively small amount of samples is enough for an accurate estimation of cardinality. For an query $Q$ whose result is $T$, let the real cardinality of $T$ be $\abs{T}$ and the estimated one be $\tilde{\abs{T}}$. Let $D = \frac{max(\tilde{\abs{T}}, \abs{T})}{min(\tilde{\abs{T}}, \abs{T})}$ be an indicator that measures their relative difference, which means the close $D$ is to $1$, the better. We conduct experiments on test-cases that consist of dataset $\LJ$ and query $Q_4, Q_5, Q_6$. For each test-case, we vary the numbers of samples from $2*10^2$ to $10^7$ and plot the maximum relative difference of all estimated cardinality and the aggregated sampling time.  The results are shown in Fig.~\ref{fig:cost_accuracy}. We can see that after the sampling budget is increased beyond $10^4$, the maximum relative difference converges to $1$, which indicates there is almost no difference between the estimated value and real value. In terms of sampling cost, before $10^6$ sampling budget, the cost stays almost the same. This experiment confirms the efficiency and accuracy of our sampling-based cardinality estimation approach.

\stitle{The Cost and Effectiveness of Co-optimization.} In this test, we show that co-optimization can effectively trading the computation with communication with a low query optimization cost, which includes the cost of sampling. We conduct experiment on test-cases that consist of datasets $\AS, \LJ, \OK$ and queries $Q_4, Q_5, Q_6$, and measures the cost of \kw{Optimization}, \kw{Pre-Computing}, \kw{Communication}, \kw{Computation} and \kw{Total}. The results are shown in Table~\ref{tab:co-opt_effectiveness_1}-Table~\ref{tab:co-opt_effectiveness_3}. From them, we can see that on almost all test-cases, when \kw{Co-Optimization} strategy is used,  with a mildly increased \kw{Pre-Computing} and \kw{Communication} cost, the \kw{Computation} cost is drastically reduced. Also, there are test-cases such as $(\OK, Q_6)$, whose \kw{Communication} cost decreases as well. The reason is that introducing pre-computed relation increases the size of the input database, but also changes the query itself and alters share $p$ of \hcube, which could result in smaller \kw{Communication} cost. From Table~\ref{tab:co-opt_effectiveness_1}-Table~\ref{tab:co-opt_effectiveness_3}, it also can be seen that although \kw{Optimization} cost of \kw{Co-Optimization} strategy is consistently larger than \kw{Optimization} cost of \kw{Communication-First\ Optimization} strategy, it is still small compared to the total cost. This experiment confirms the effectiveness of \kw{Co-Optimization} strategy and relatively low query \kw{Optimization} cost of \kw{Co-Optimization} strategy. 

\begin{table*}[t!]\centering
\resizebox{0.98\textwidth}{!}{%
\begin{tabular}{@{}llp{0cm}lp{0cm}lp{0cm}lp{0cm}lp{0cm}lp{0cm}lp{0cm}lp{0cm}lp{0cm}l@{}}\toprule
& \multicolumn{9}{c}{\kw{Co-Optimization (sec)}} & \phantom{abc}& \multicolumn{7}{c}{\kw{Communication-First \ Optimization (sec)}}  \\ \cmidrule{2-10} \cmidrule{12-18}
& \kw{Optimization} & \phantom{abc}& \kw{Pre-Computing} &
\phantom{abc} & \kw{Communication} & \phantom{abc} & \kw{Computation} & \phantom{abc}& \kw{Total} & \phantom{abc}& \kw{Optimization} &
\phantom{abc} & \kw{Communication} & \phantom{abc} & \kw{Computation} & \phantom{abc} & \kw{Total} \\ \midrule
$Q_4$  & $107$ && $12$ && $66$  && $1276$ && $1461$ && $3$ && $21$  && $>43200$ && $>43200$\\
$Q_5$  & $90$ && $24$ && $50$  && $907$ && $1071$ && $4$ && $36$  && $>43200$ && $>43200$\\
$Q_6$  & $63$ && $12$ && $19$  && $18$ && $112$ && $4$ && $47$  && $30426$ && $30477$ \\
\bottomrule
\end{tabular}
}

\caption{The comparison between co-optimization and communication-first optimization strategy in \AS dataset\label{tab:co-opt_effectiveness_1}}
\vspace{0.2cm}
\resizebox{0.98\textwidth}{!}{%
\begin{tabular}{@{}llp{0cm}lp{0cm}lp{0cm}lp{0cm}lp{0cm}lp{0cm}lp{0cm}lp{0cm}lp{0cm}l@{}}\toprule
& \multicolumn{9}{c}{\kw{Co-Optimization (sec)}} & \phantom{abc}& \multicolumn{7}{c}{\kw{Communication-First \ Optimization (sec)}}  \\ \cmidrule{2-10} \cmidrule{12-18}
& \kw{Optimization} & \phantom{abc}& \kw{Pre-Computing} &
\phantom{abc} & \kw{Communication} & \phantom{abc} & \kw{Computation} & \phantom{abc}& \kw{Total} & \phantom{abc}& \kw{Optimization} &
\phantom{abc} & \kw{Communication} & \phantom{abc} & \kw{Computation} & \phantom{abc} & \kw{Total} \\ \midrule
$Q_4$  & $106$ && $22$ && $132$  && $1282$ && $1542$ && $8$ && $62$  && $>43200$ && $>43200$\\
$Q_5$  & $132$ && $44$ && $103$  && $222$ && $501$ && $9$ && $112$  && $>43200$ && $>43200$\\
$Q_6$  & $105$ && $22$ && $147$  && $350$ && $624$ && $12$ && $204$  && $>43200$ && $>43200$\\
\bottomrule
\end{tabular}
}

\caption{The comparison between co-optimization and communication-first optimization strategy in \LJ dataset\label{tab:co-opt_effectiveness_2}}
\vspace{0.2cm}
\resizebox{0.98\textwidth}{!}{%
\begin{tabular}{@{}llp{0cm}lp{0cm}lp{0cm}lp{0cm}lp{0cm}lp{0cm}lp{0cm}lp{0cm}lp{0cm}l@{}}\toprule
& \multicolumn{9}{c}{\kw{Co-Optimization (sec)}} & \phantom{abc}& \multicolumn{7}{c}{\kw{Communication-First \ Optimization (sec)}}  \\ \cmidrule{2-10} \cmidrule{12-18}
& \kw{Optimization} & \phantom{abc}& \kw{Pre-Computing} &
\phantom{abc} & \kw{Communication} & \phantom{abc} & \kw{Computation} & \phantom{abc}& \kw{Total} & \phantom{abc}& \kw{Optimization} &
\phantom{abc} & \kw{Communication} & \phantom{abc} & \kw{Computation} & \phantom{abc} & \kw{Total} \\ \midrule
$Q_4$  & $218$ && $71$ && $712$  && $13214$ && $14215$ && $37$ && $1050$  && $>43200$ && $>43200$\\
$Q_5$  & $265$ && $142$ && $422$  && $877$ && $1706$ && $46$ && $1566$  && $>43200$ && $>43200$\\
$Q_6$  & $278$ && $71$ && $1189$  && $516$ && $2054$ && $42$ && $2067$  && $>43200$ && $>43200$\\
\bottomrule
\end{tabular}
}

\caption{The comparison between co-optimization and communication-first optimization strategy in \OK dataset \label{tab:co-opt_effectiveness_3}}


\vspace*{-0.15cm}
\end{table*}

\begin{figure}
	[tbp] \centering 
	\begin{center}

		\hspace*{-0.8cm} 
		\begin{tabular}
			[t]{c} \subfigure[$Q_1$] { 
			\includegraphics[width=0.375\columnwidth]{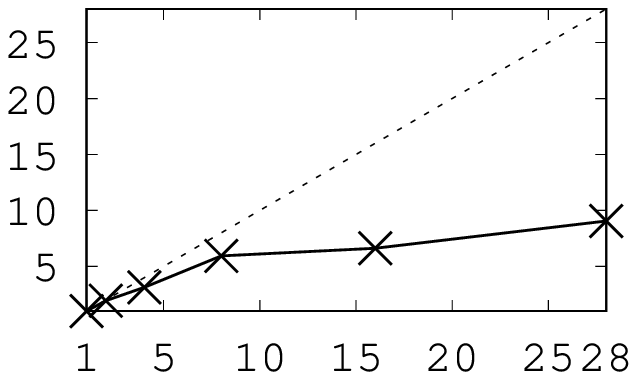} } \hspace{-0.7cm} \subfigure[$Q_2$] { 
			\includegraphics[width=0.375\columnwidth]{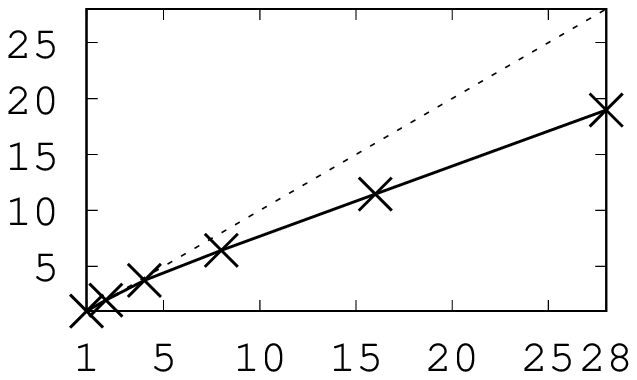} } \hspace{-0.7cm} \subfigure[$Q_3$] { 
			\includegraphics[width=0.375\columnwidth]{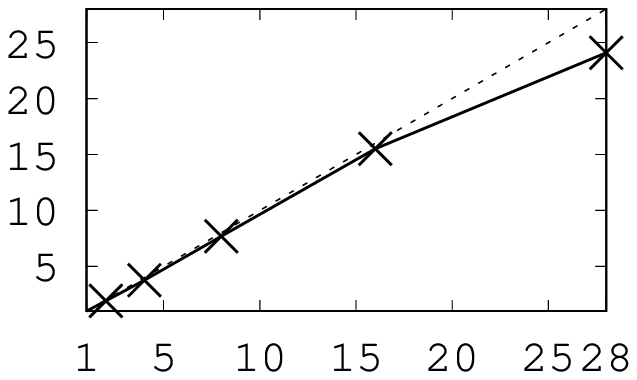} } 
		\end{tabular}
		
		\hspace*{-0.8cm} 
		\begin{tabular}
			[t]{c} \subfigure[$Q_4$] { 
			\includegraphics[width=0.375\columnwidth]{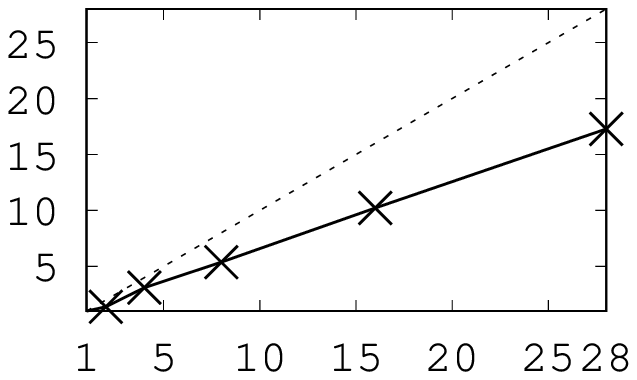} } \hspace{-0.7cm} \subfigure[$Q_5$] { 
			\includegraphics[width=0.375\columnwidth]{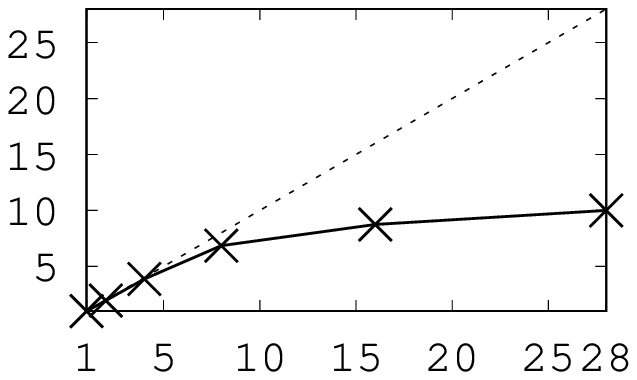} } \hspace{-0.7cm} \subfigure[$Q_6$] { 
			\includegraphics[width=0.375\columnwidth]{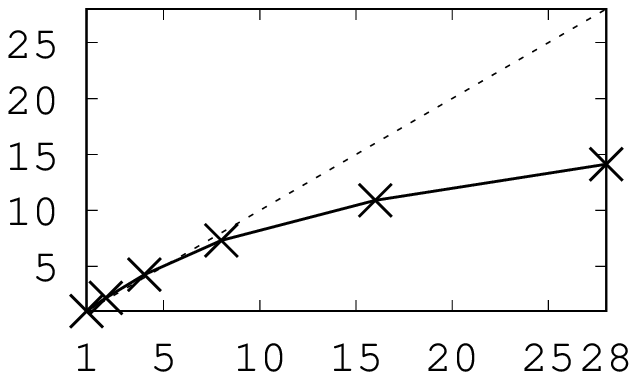} } 
		\end{tabular}
	\end{center}
	\vspace{-0.2cm}
	
	\caption{Speed-up factor of \apj under difference workers under $1$ to $28$ workers \label{fig:Vcore}.} \vspace{-0.6cm} 
\end{figure}

\stitle{Scalability.}  In Fig~\ref{fig:Vcore}, we show the speedup of our system when varying the number of workers of Spark from $1$ to $28$ on test-cases that consist of $\LJ$, and all queries. It can be seen that our system has a near-linear speed up on query $Q_2, Q_3, Q_4, Q_6$. For query $Q_1$, the scalability is limited as it is a rather simple query, and the overhead of the systems gradually becomes the dominating cost. For query $Q_5$, its limited scalability is due to the skewness, where the ``last straggler" effect plays a bigger role in determining the elapsed time.

%
%
%
%
%

\begin{figure*}
	[tbp] \centering 
	\begin{center}
		\begin{tabular}
			[t]{c} \subfigure { 
			\includegraphics[width=1.4\columnwidth]{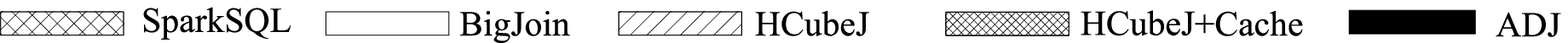} } \addtocounter{subfigure}{-1} 
		\end{tabular}
		
		\begin{tabular}
			[t]{c} \hspace*{-0.5cm} \subfigure[$Q_1$] { 
			\includegraphics[width=0.70\columnwidth]{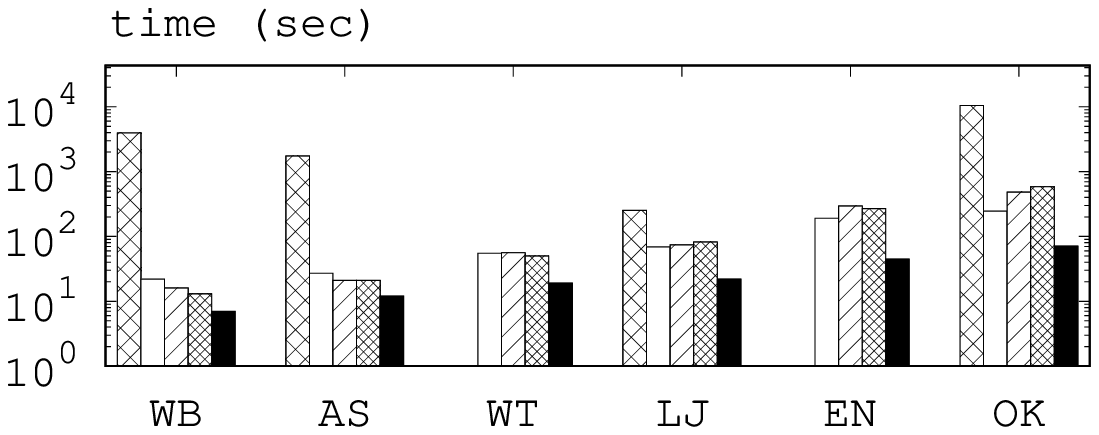} } \hspace*{-0.5cm} \subfigure[$Q_2$] { 
			\includegraphics[width=0.70\columnwidth]{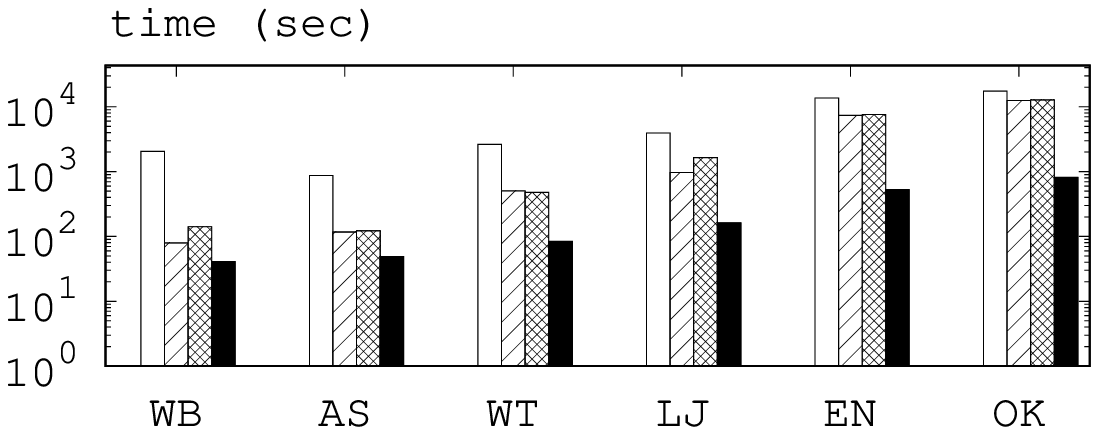} } 
			
			\hspace*{-0.5cm} \subfigure[$Q_3$] { 
			\includegraphics[width=0.70\columnwidth]{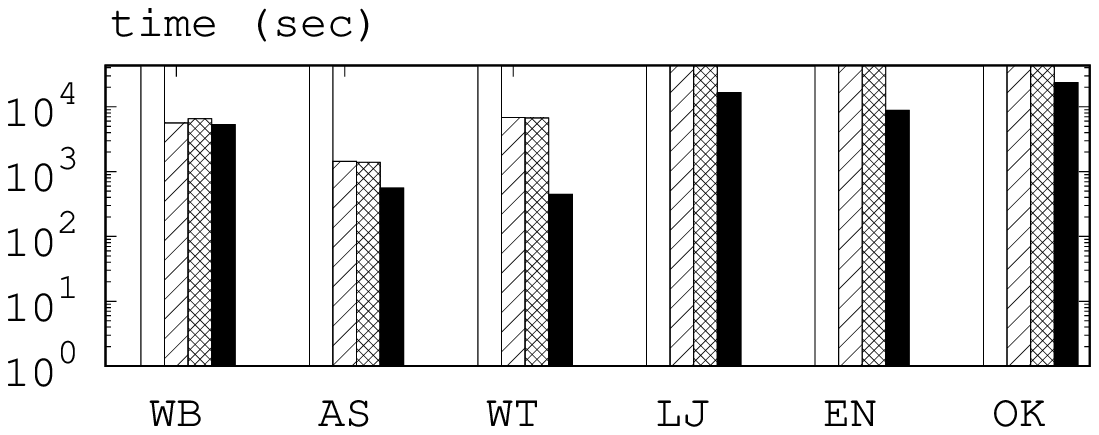} } 
		\end{tabular}
		
		\vspace*{-0.2cm} 
		\begin{tabular}
			[t]{c} \hspace*{-0.5cm} \subfigure[AS] { 
			\includegraphics[width=0.70\columnwidth]{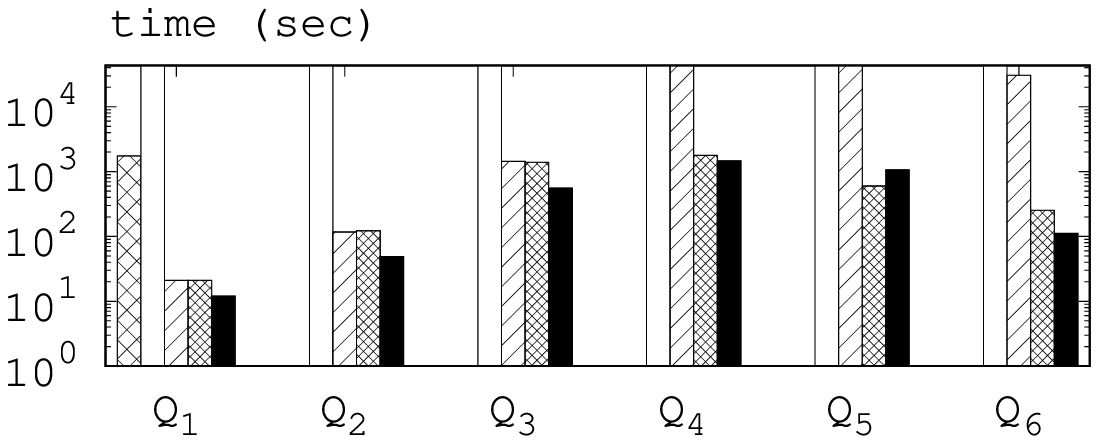} } \hspace*{-0.5cm} \subfigure[LJ] { 
			\includegraphics[width=0.70\columnwidth]{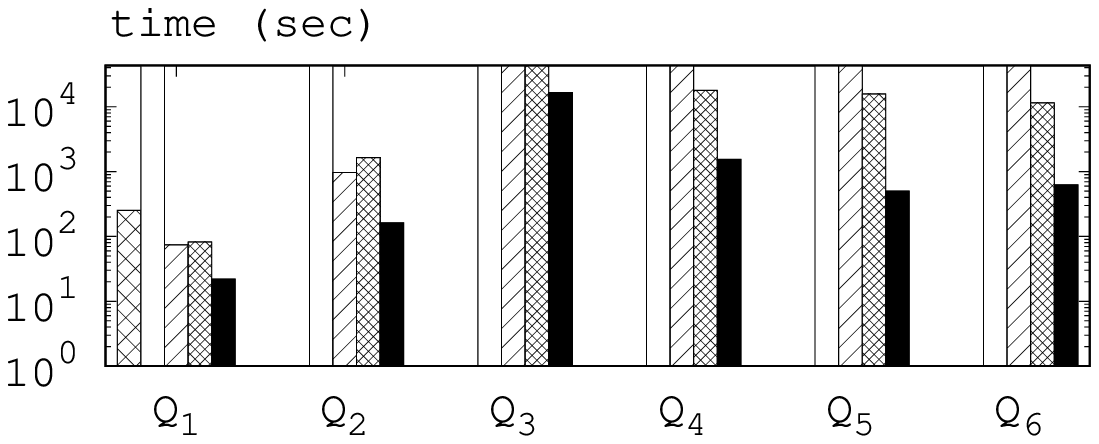} } \hspace*{-0.5cm} \subfigure[OK] { 
			\includegraphics[width=0.70\columnwidth]{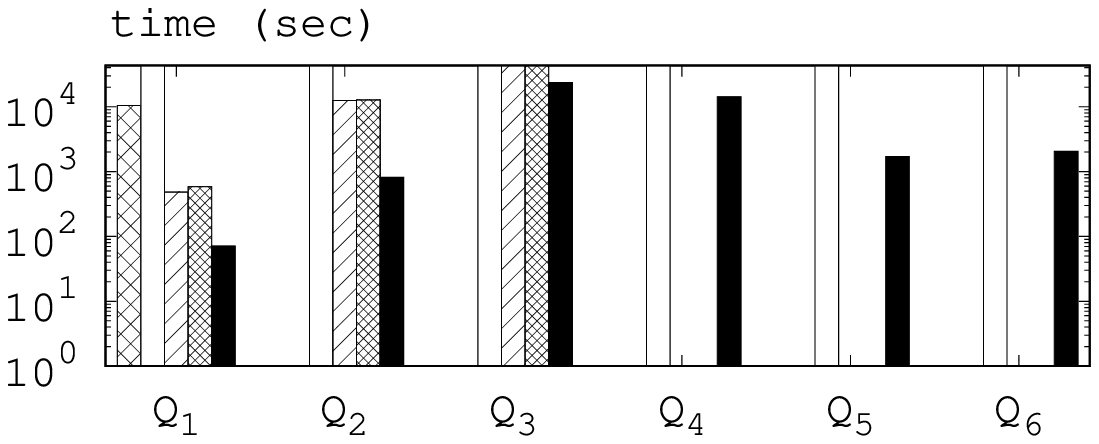} } 
		\end{tabular}
	\end{center}
	
	\vspace*{-0.2cm} 
	\caption{Comparison of methods by varying datasets or queries \label{fig:exp_varying}} \vspace*{-0.4cm} 
\end{figure*}

\subsection{Comparison with Other Join Approaches} 

In this section, we compare \apj against state-of-the-art methods. 

\stitle{Varying Dataset.} In this test, we compare each method on test-cases where the queries are fixed to $Q_1, Q_2, Q_3$. The results are shown in Fig.~\ref{fig:exp_varying} (a)-(c). It can be seen that multi-round methods \kw{SparkSQL} and \kw{BigJoin} fail on many of the queries due to overwhelming intermediate results, while one-round methods successfully tackle most of the queries as the shuffling of intermediate results are avoided. Also, \kw{BigJoin} is better than \kw{SparkSQL} as the worst-case optimal join \lf it parallelizes generates less intermediate tuples. Also, it can be seen that with the increase the input database size, \hcubej, \kw{HCubeJ+Cache}, spent more portion of time on \hcube, and on test-case $(\LJ, Q_3)$, $(\EN, Q_3)$, $(\OK, Q_3)$, they have a difficult time shuffling the tuples using original \hcube implementation. In comparison, \apj can successfully process all test-cases and spent significantly less time when shuffling the relations on test-cases that involve complex queries such as $Q_3$ or large dataset $\EN, \OK$.

\stitle{Varying Query.} In this test, we compare each method on test-cases where the datasets are fixed to $\AS, \LJ, \OK$. The results are shown in Fig.~\ref{fig:exp_varying} (d)-(e). For \kw{SparkSQL}, it can only handle $Q_1$ and failed on all other queries due to overwhelming intermediate results. And, \kw{BigJoin} can only handle $Q_1$ and $Q_2$. For $Q_1 - Q_3$, \hcubej and \kw{HCubeJ+Cache} performs similarly, and \apj has a large lead due to the optimized \hcube. For $Q_4 - Q_6$, \kw{HCubeJ+Cache} performs better than \hcubej, and \kw{HCubeJ+Cache} has similar performance to \apj on dataset $\AS$ as $\AS$ is relatively small and there is abundant remaining memory on each server to use for caching. On $\LJ$ dataset, \kw{HCubeJ+Cache} is significantly outperformed by \apj, as \kw{HCubeJ+Cache} is a method that prioritizes communication cost over computation cost, and uses up all memory for shuffling and storing the tuples during \hcube, which leaves little memory for caching. On $\OK$ dataset, both \hcubej and \kw{HCubeJ} \kw{+Cache} failed, as the original \hcube implementation shuffles too many tuples, which causes memory-overflow. It can be seen that in almost all test-case \apj can effectively balance the computation cost and communication cost by adopting a co-optimization strategy.

\section{Conclusion} \label{sec:conclude} 
This paper studies the problem of co-optimize communication and computation cost in a one-round multi-way join evaluation and proposes a prototype system \apj for processing complex join queries. To find an effective query plan in a huge search space in terms of total cost, this paper study how to restrict the search space based on an optimal hypertree ${\cal T}$ and how to explore cost-effective query plans based on hypertree ${\cal T}$. Extensive experiments have shown the effectiveness of various optimization proposed in \apj. We shall explore co-optimize computation, pre-computing, and communication for a query that consists of selection, projection, and join. 

\section*{Acknowledgement}
{This work is supported by the Research Grants Council of Hong Kong, China under No. 14203618, No. 14202919 and No. 14205520, No. 14205617, No. 14205618, and NSFC Grant No. U1936205.
}

{
\bibliographystyle{ieeetr}
\bibliography{ADJ}
}

\end{document}